\documentclass[conference]{IEEEtran}

\usepackage{subfigure}
\usepackage{cite}
\usepackage{booktabs}
\usepackage{amsmath,amssymb,amsfonts}
\usepackage{algorithmic}
\usepackage{graphicx}
\usepackage{textcomp}
\usepackage{xcolor}
\def\BibTeX{{\rm B\kern-.05em{\sc i\kern-.025em b}\kern-.08em
    T\kern-.1667em\lower.7ex\hbox{E}\kern-.125emX}}

\usepackage{booktabs}
\usepackage{times}
\usepackage{adjustbox}
\usepackage{multicol}

\begin{document}

\title{A Cognitive Network Architecture for Vehicle-to-Network (V2N) Communications over Smart Meters for URLLC \\

\thanks{}
}

\title{A Cognitive Network Architecture for Vehicle-to-Network (V2N) Communications over Smart Meters for URLLC}
\author{\IEEEauthorblockN{Shoaib Ahmed\textsuperscript{1}, Sayonto Khan\textsuperscript{2}, Kumudu S. Munasinghe\textsuperscript{3} and Md. Farhad Hossain\textsuperscript{4}}
\IEEEauthorblockA{\textit{\textsuperscript{1,2,4}Department of Electrical and Electronic Engineering}
\\ \textit{Bangladesh University of Engineering and Technology, Dhaka-1205, Bangladesh} \\ \textit{\textsuperscript{3}School of Information Technology and Systems University of Canberra, Canberra, ACT 2617, Australia}\\
shoaibahmedshb@gmail.com\textsuperscript{1}, sayontokhan17103@gmail.com\textsuperscript{2}, kumudu.munasinghe@canberra.edu.au\textsuperscript{3}, and \\ mfarhadhossain@eee.buet.ac.bd\textsuperscript{4}}
}

\maketitle

\begin{abstract}
With the rapid advancement of smart city infrastructure, vehicle-to-network (V2N) communication has emerged as a crucial technology to enable intelligent transportation systems (ITS). The investigation of new methods to improve V2N communications is sparked by the growing need for high-speed and dependable communications in vehicular networks. To achieve ultra-reliable low latency communication (URLLC) for V2N scenarios, we propose a smart meter (SM)-based cognitive network (CN) architecture for V2N communications. Our scheme makes use of SMs' available underutilized time resources to let them serve as distributed access points (APs) for V2N communications to increase reliability and decrease latency. We propose and investigate two algorithms for efficiently associating vehicles with the appropriate SMs. Extensive simulations are carried out for comprehensive performance evaluation of our proposed architecture and algorithms under diverse system scenarios. Performance is investigated with particular emphasis on communication latency and reliability, which are also compared with the conventional base station (BS)-based V2N architecture for further validation. The results highlight the value of incorporating SMs into the current infrastructure and open the door for future ITSs to utilize more effective and dependable V2N communications.
\end{abstract}


\section{Introduction}

\label{Introduction}

The concept of smart city is emerging as a reality with the development of information and communication technology (ICT), including internet-of-things (IoT), cloud computing, artificial intelligence (AI), big data, and fifth-generation (5G) cellular networks \cite{b1}. A study presented in \cite{b2} has identified six smart characteristics of smart cities to be considered: economy, governance, environment, people, transportation, and living. Over the past two decades, the use of communication technology in vehicles gradually emerges as a rising trend to ensure smart transportation in a smart city. In 1999, the U.S. Federal Communications Commission (FCC) allocated $75\,\mathrm{MHz} $ spectrum in the $5.9\, \mathrm{GHz}$ band for vehicular communications \cite{b3}. The allocation triggered numerous research activities worldwide to develop and deploy ITS for various applications such as traffic management, safety warnings, and autonomous driving \cite{b4}. Vehicle-to-everything (V2X) communication includes vehicle-to-vehicle (V2V), vehicle-to-network (V2N), vehicle-to-road side unit (V2R), and vehicle-to-pedestrian (V2P). Data exchange between vehicles and the network infrastructure is defined as V2N communication. V2N communication demands seamless connectivity and real-time data exchange between vehicles and the underlying network infrastructure, which is quite difficult to achieve. URLLC is one of the service categories of 5G as well as of the emerging sixth-generation (6G) cellular networks for successfully delivering packets with stringent requirements, particularly regarding availability, latency, and reliability \cite{b5}. Low latency makes it possible to send and receive data quickly, while reliable communication ensures that data is accurately sent without loss or damage. So, to enhance the capability of V2N communications to implement the applications of ITS, URLLC for V2N communications is mandatory\cite{b6}. An end-to-end (E2E) latency of $1ms$ and reliability above $99\%$ is required for ITS \cite{b5,b7}.

To achieve these requirements, we propose a new architecture considering SMs of smart grid system as the distributed APs for utilizing the available unused time resources they offer. The motivation of this work is to exploit the availability of SMs for achieving reliability and latency criteria for V2N URLLC. SMs are prevalent in utility networks and are regularly used to collect data regarding the consumption of electrical energy \cite{b8}. However, during operational cycles, SMs have abundant free time slots when the assigned electromagnetic spectrum to SM remains underutilized. This spectrum underutilization could be employed for other tasks, such as V2N communications \cite{b9}. 

On the other hand, cognitive networks (CNs) are self-aware in the sense that they can make configuration decisions in the context of a specific environment and uses the methodology of understanding by-building to learn from the environment and adapt its internal states to ensure the efficient utilization of assigned spectrum \cite{b10}. As the SMs have underutilized spectrum in their operational cycles, spectrum utilization could be improved by enabling a secondary user to access the spectrum. As we integrate SMs into the V2N communication architecture, now the vehicles are allowed to access the underutilized spectrum of SMs, and we consider the proposed network architecture as a cognitive one. A CN, with its adaptive resource allocation, adaptability, and capacity for intelligent decision-making, outperforms conventional methods in terms of performance \cite{b11}.

To this end, by combining the facts that ITS demands URLLC for seamless communications and the SMs have underutilized spectrum, which could be accessed by vehicles in a cognitive manner, this paper proposes a CN architecture for V2N communications over SMs for URLLC with the following contributions.

\begin{itemize}
    \item We propose a new network architecture using SMs as distributed APs of communications, ensuring increased reliability and reduced latency for enhancing V2N communications. Both single-hop vehicle-to-smart meter (V2SM) communications and multi-hop V2SM communications over other vehicles for vehicle-SM associations are considered in the proposed system model.
    
    \item Two distinct algorithms, namely, the maximum SNR (MaxSNR) algorithm and the minimum distance (MinDis) algorithm, are proposed for vehicle-SM associations. The MaxSNR algorithm aims to maximize the SNR of each link for both V2SM and V2V communications using an iterative process. On the contrary, the MinDis algorithm searches for the SM or the vehicle having minimum distance for V2SM or V2V associations. 
    
    \item We conduct extensive investigations by developing a MATLAB simulation platform for determining the viability and acceptance of our proposed system model and algorithms.
    
    \item Various performance metrics, including reliability, latency, and throughput are evaluated under various system settings for validating the effectiveness of the proposed architecture for V2N URLLC operations.
    
    \item The trivial BS-based V2N system model is also simulated and compared to determine the superiority of our proposed system model.
    
\end{itemize}

The organization of the rest of the paper is as follows: Section \ref{Rationality of the proposed architecture} addresses the rationality of the proposed architecture with an analysis of the architecture of smart grid communications, CN, and BS-based V2N system. Section \ref{Proposed system model} comprehensively discusses the proposed system model and the wireless channel model. Section \ref{Performance metrics} defines performance metrics, namely, SNR, latency, reliability, and throughput, which are used for evaluating system performance, while section \ref{Proposed algorithms} presents the proposed algorithms. Then, section \ref{Results and discussions} introduces the simulation environments and analyzes the results found through extensive simulations. Finally, section \ref{Conclusions} concludes the paper by summarizing the essential findings and highlighting the research issues that merit attention in future developments.

\section{Rationality of the proposed architecture}
\label{Rationality of the proposed architecture}

\subsection{Smart grid communication and available resources of SMs}

Traditionally, smart grid communication refers to bidirectional data and energy flow depending on its intelligence, energy, and infrastructure. The conventional unidirectional grid was based on providing the generated power to load sites and consumer ends over transmission and distribution lines. However, the smart grid allows load sites or customers to generate energy and supply it to the utility grid using micro-generation sources, enabling bidirectional energy flow \cite{b12}. The communication architecture of the smart grid is defined by the IEEE 2030-2011 standard, which is essential to understanding the applications and infrastructures in a hierarchical arrangement \cite{b13}. The smart grid server utilizes automatic meter reading (AMR) technique that uses necessary communication technology to gather meter readings, events, and alarm data from the meters. Several published standards are already available in this field; among them, ANSI C12.1-2008, IEEE 1377, and IEC 61968-9 are the prominent standards that provide specifications for the communication syntax for data exchange between the end device and the utility server \cite{b14,b15,b16}. The devices use binary codes and extensible markup language (XML) codes.

\begin{figure}
\centering
\centerline{\includegraphics[width=\columnwidth]{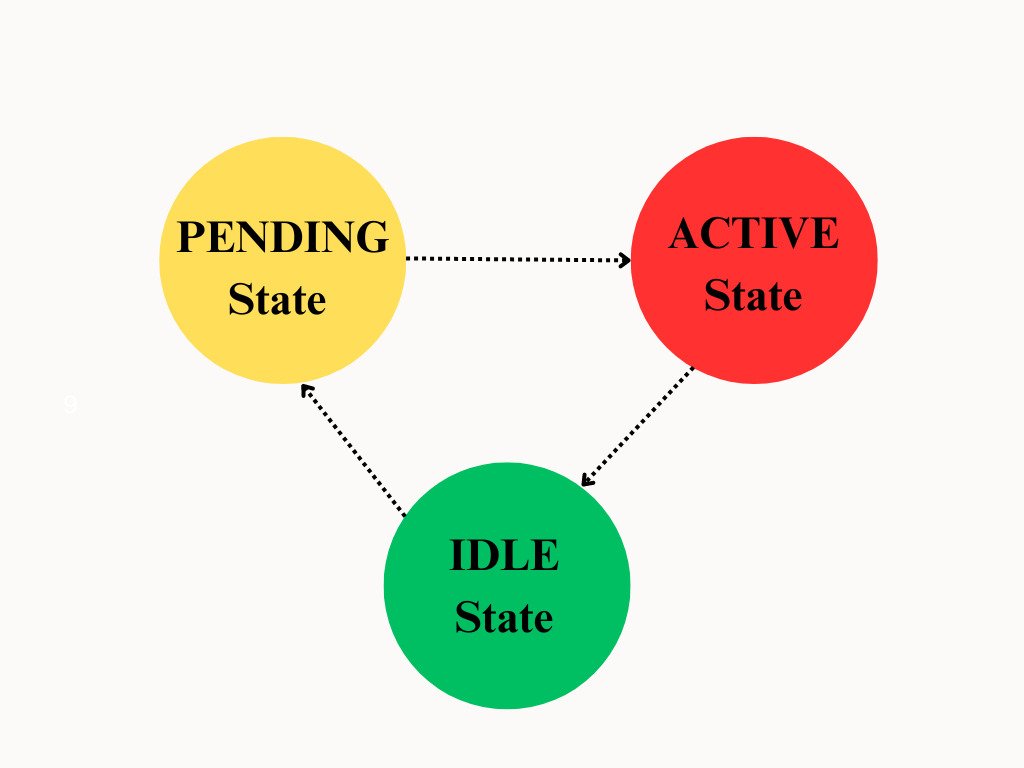}}
\caption{State transmission diagram for DRAS client in smart grid.}
\label{fig: State Transmission Diagram for DRAS Client}
\end{figure}

A typical meter reading report is $100$ to $200$ bytes according to the message format specified \cite{b17}. Thus, the client end and utility provider end can communicate about generated power demand and alarm events via demand response (DR). It enables the utility operator to optimally balance generated power and consumption by offering dynamic pricing or implementing various load control programs. For DR applications, the open automated demand response (OpenADR) is the pioneering standard that provides specifications and guidance for the automation of DR programs \cite{b18}. Upon receiving the price signals from the demand response automation server (DRAS) in DR applications, the customer takes the necessary steps to regulate energy usage. The client-end DRAS uses a state transition diagram with three states: idle state, pending state, and active state, as shown in Figure \ref{fig: State Transmission Diagram for DRAS Client} \cite{b19}. In any event, the client's DRAS conveys a message, and the demand remains pending. The time between staying idle and pending is the event issue time. Finally, the event is answered and remains active until the event's end time. Then, the devices stay idle, and several studies show the idle state remains around 15 minutes or less when the assigned electromagnetic spectrum to SM remains largely idle \cite{b17}. This phenomenon gives a substantial opportunity for vehicles to be eligible for access to the network through SMs.

\subsection{Cognitive network}
A CN supports massive connectivity through efficient spectrum management while overseeing its environment variables and making effective decisions. Moreover, CNs employ intelligent decision-making ability to determine the best access strategies. Also, if provided with necessary information variables, CNs can learn and adapt to the ever-changing environment and user behaviors \cite{b11,b20,b21}. Spectrum hole (SH) is defined as a band of frequencies assigned to a primary user (PU) that remains underutilized and that can be utilized by unlicensed cognitive radio (CR) users (i.e., also termed as the secondary user (SU)), which is an essential resource for CN systems \cite{b22,b23}. CN offers a way of efficient use of the radio spectrum by exploiting the existence of SH. A CR user monitors the available spectrum bands and detects SH. Based on the spectrum availability and internal policy, CR users can allocate a channel from SH. Moreover, there is a possibility of multiple users trying to access the same SH, so network access is coordinated to prevent multiple users from colliding in overlapping portions of the spectrum \cite{b24,b25}.   

 There have been numerous studies conducted about V2N and vehicle-to-infrastructure (V2I) communications in the recent past, and the inclusion of different roadside units (RSU) into the network as infrastructure is also introduced by researchers \cite{b26}. This approach of including RSU as network infrastructure motivates us to propose SM as RSU, where SM has an underutilized spectrum, as mentioned in the previous subsection, that can be utilized by SUs of a CN. For our proposed cognitive architecture, we consider vehicles as SUs with the capability to exploit the SH offered by SMs. As CN provides efficient spectrum utilization, energy efficiency, and better E2E performance than a non-CN, the proposed CN architecture is expected to support improved V2N communications \cite{b21,b27,b28}.

\subsection{Conventional BS-based V2N system}
\begin{figure}
\centering
\centerline{\includegraphics[width=\columnwidth]{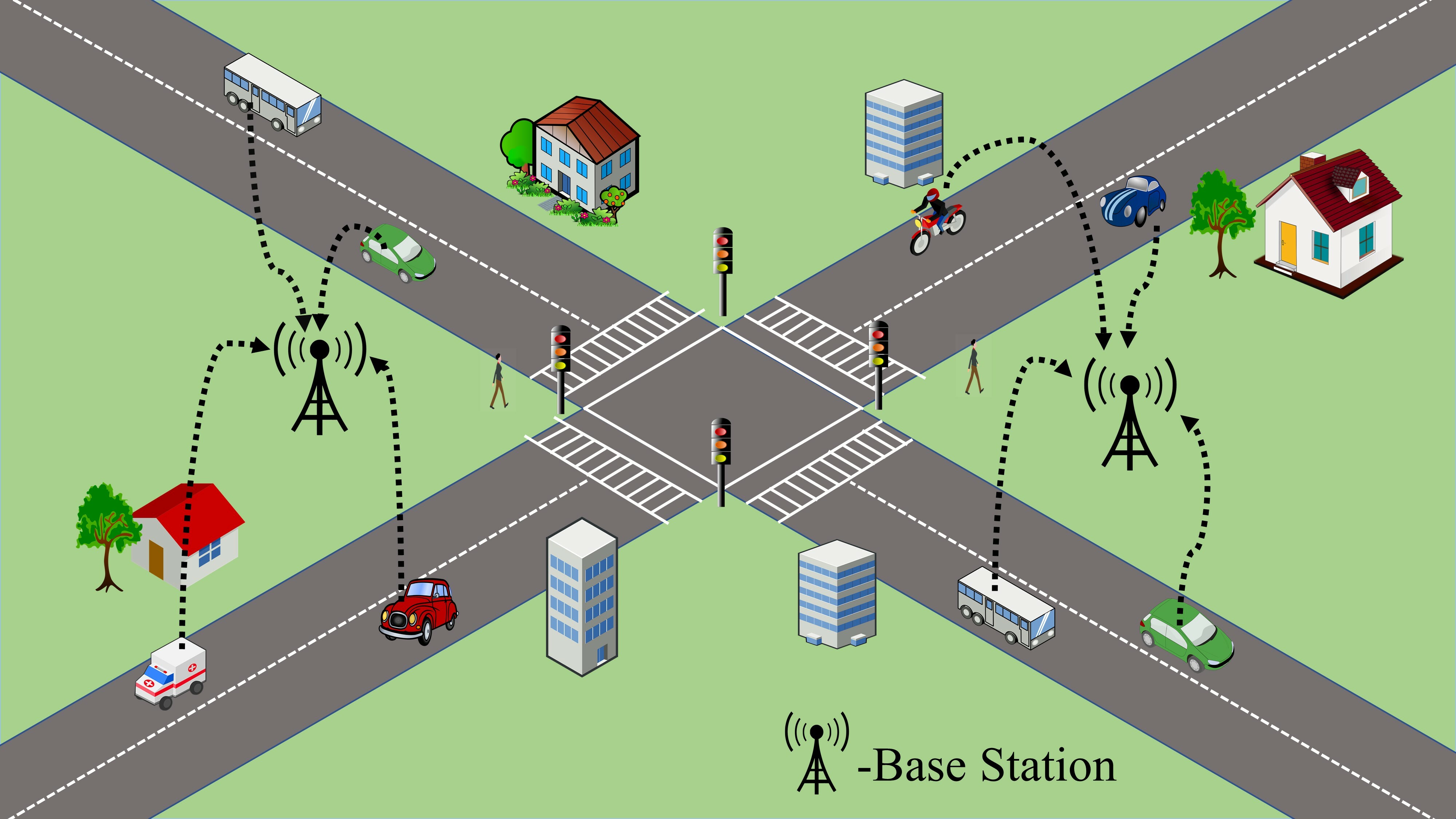}}
\caption{Conventional BS-based V2N system model.}
\label{Conventional BS-based V2N system model}
\end{figure}

The BS-based communication system is the conventional method for V2N communications, where a BS is considered as a central hub for receiving and transmitting data from both vehicles and network infrastructure \cite{b29}, as shown in Figure \ref{Conventional BS-based V2N system model}. The BS handles tasks such as channel allocation, signal processing, and routing, ensuring seamless connectivity and data exchange between vehicles and the network. The number of required BSs is determined based on the dimensions and coverage requirements of the area. A single BS is deployed within a geographical area typically covering $1000\,m\times1000\,m$ to adapt to the communication range of vehicles \cite{b30}. BS-based communications cover various ranges, including V2V, V2I, V2N, and V2P communications. Vehicles within proximity of each other can exchange information either directly or with the help of BS, whereas BS also exchanges unicast information with vehicles. LTE-V2X standard is a model developed for compatible vehicular communication systems with 4G LTE BSs \cite{b31}. 3rd generation partnership project (3GPP) provides architectural support for LTE-based V2X. 3GPP suggests two models for vehicles to BS association, i.e., PC5 and LTE-Uu. The PC5-based model is followed for the case of V2V communications. On the contrary, LTE-Uu connects vehicles to BS for the provision of V2N services \cite{b32}.

While BS-based communication offers a foundation for V2N communication, this system has some significant drawbacks. Firstly, the coverage area of a vehicle is limited, and the deployment of BS at a large scale isn't feasible, so a possible reduction in reliability is inevitable. Moreover, LTE has a centralized nature, due to which vehicular data needs to be passed through the BS and the core network, which can raise the issue of latency, considering the scenario of increased vehicle density. It has also been shown that existing beaconing capabilities of LTE-operated BS to support vehicular safety applications are poor \cite{b33}.

As it stands now, SMs have an underutilized spectrum that vehicles could access as CR SUs in a CN. CN provides additional advantages compared to a non-CN and maintains improved reliability and latency, which is essential for URLLC. In addition, the typically used BS-based communication model of V2N communication has some drawbacks. By undertaking those issues, we consider SMs as distributed APs for V2N connectivity and propose a new SM-based CN architecture for V2N communications which is inferred to transgress the BS-based model by performance.

\section{Proposed system model}
\label{Proposed system model}

\subsection{Proposed SM-based architecture}

\begin{figure}[htbp]
\centering
\centerline{\includegraphics[width=\columnwidth]{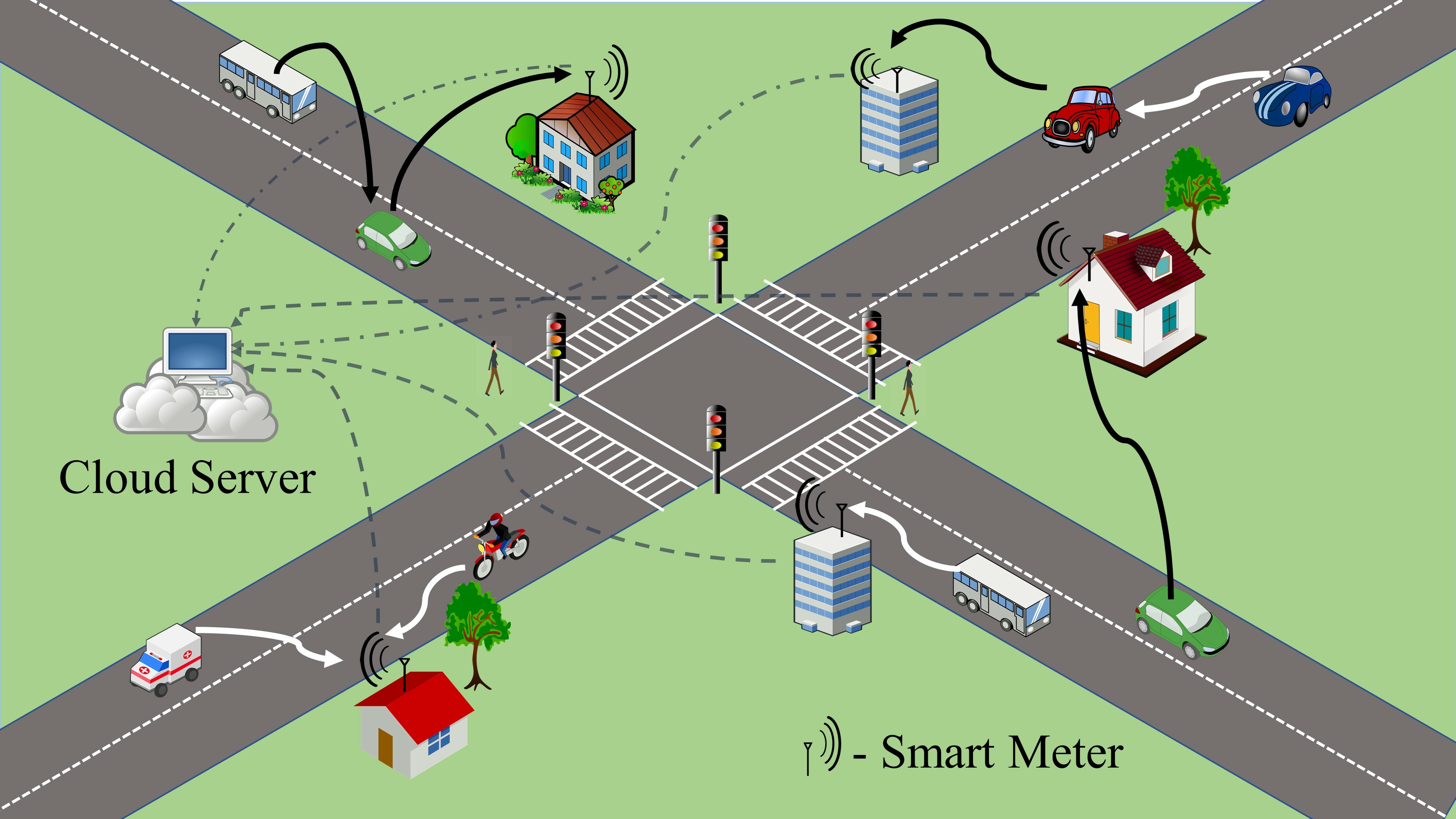}}
\caption{Proposed SM-based V2N system model.}
\label{SM-based system model.}
\end{figure}

Each house in a smart city is assumed to have an SM, resulting in a network infrastructure that aligns with the density of residential units. The availability of this abundant number of SMs as network APs drastically increases the resources for vehicles to connect with the network. We consider a suburban area as shown in Figure \ref{SM-based system model.} with $N$ SMs and $M$ vehicles. To establish a connection with the network, $i^{th}$ vehicle can directly communicate with a nearby SM or opt for a multi-hop relay approach via other vehicles based on a predefined algorithm as presented in section \ref{Proposed algorithms}. Vehicles are equipped with communication modules like transmitting and receiving antennas and continuously assess the information about SNR and distance of nearby SMs and vehicles. Based on the deployed algorithm, a vehicle selects the appropriate SM through which a vehicle will ensure connectivity with the network. Moreover, all SMs within the system are connected to a smart grid cloud server, either using a wireless link or through a high-speed gigabit passive optical network (GPON) connection. To ensure seamless communication, we adopt the IEEE 802.11p protocol as the wireless link standard for both V2V and V2SM communications \cite{b34}. 

In comparison with a BS-based model, the SM-based communication system has the potential to offer several advantages. Firstly, the distribution of SMs throughout the suburban area ensures denser APs for communications within a short distance that reduces signal attenuation. So an improved signal quality with low latency and enhanced reliability for V2N communications can be ensured, leading to URLLC. Secondly, as we use the existing infrastructure of SMs, it significantly reduces the need to establish additional infrastructure, thereby minimizing costs and resource requirements.

\subsection{Wireless channel model}

We consider the Urban Micro (UMi) street canyon environment modeled by the 3GPP \cite{b35}. The path-loss model is given below.

\begin{align}
&PL_{UMi-LOS} = \begin{cases}PL_{1} &  \quad if  \quad 10m  \leq d_{2D}  \leq d_{BP}^{\prime} \\PL_{2} &  \quad if  \quad  d_{BP}^{\prime}  \leq d_{2D}  \leq 5km \end{cases} \\
&PL_{1} = 32.4 + 21log_{10}(d_{3D}) + 20log_{10}(f_{c}) \label{2}\\
&PL_{2} = 32.4 + 40log_{10}(d_{3D}) + 20log_{10}(f_{c}) \nonumber\\
&\quad\quad\quad- 9.5log_{10}({d_{BP}^{\prime}}^{2} + (h_{BS} - h_{UT})^{2}) \label{3}\\
&PL_{UMi-NLOS} = max(PL_{UMi-LOS},PL_{UMi-NLOS}^{\prime}) \nonumber\\
&\quad\quad\quad  for \quad 10m  \leq d_{2D}  \leq 5km \\ 
&PL_{UMi-NLOS}^{\prime} = 35.3log_{10}(d_{3D}) + 22.4 \nonumber\\ 
&\quad\quad\quad + 21.3log_{10}(f_{c}) - 0.3(h_{UT} - 1.5) \label{5}\\
&PL = max(PL_{UMi-LOS},PL_{UMi-NLOS}) + N(0,\sigma_{SF}^{2}) \\
&\sigma_{SF} = \begin{cases} 4 dB & for \quad  UMi \quad  LOS \\7.82 dB &  for\quad   UMi \quad  NLOS  \end{cases} \label{shadowfading}
\end{align}
where,
\begin{description}
    \item[$PL$] = Path loss (PL)
    \item[$PL_{UMi-LOS}$] \hspace{1cm} = UMi line-of-sight (LOS) path loss 
    \item[$PL_{UMi-NLOS}$] \hspace{1.2cm} = UMi non line-of-sight (NLOS) path loss 
    \item[$d_{2D}$] = Two-dimensional distance between transmitter and receiver
    \item[$d_{3D}$] = Three-dimensional distance between transmitter and receiver
.    \item[$F_{c}$] = Center frequency/frequency
    \item[$d_{BP}^{\prime}$] = Breakpoint distance
    \item[$h_{BS}$] = Height of BS either SM/ Vehicle in our case
    \item[$h_{UT}$] = Height of user terminal (UT), vehicle in our case
    \item[$\sigma_{SF}$] = Standard deviation of log-normally distributed shadow fading (SF)
\end{description}

 Here, breakpoint distance $d_{BP}^{\prime}$ can be represented by $d_{\mathrm{BP}}^{\prime}=4 h_{\mathrm{BS}}^{\prime} h_{\mathrm{UT}}^{\prime} f_{\mathrm{c}} / c $ with $h_{BS}^{\prime} = h_{BS} - h_E $ and $h_{UT}^{\prime} = h_{UT} - h_E$, where the value of $h_E$ is typically taken as $1\,\mathrm{m}$. The above equations provide the PL model for the UMi environment in both LOS and NLOS scenarios, considering all the specified parameters. The PL in UMi is divided into two scenarios: LOS and NLOS. For the LOS case, the PL is determined based on the Euclidian distance between the transmitter and receiver, denoted as $d_{2D}$. If $10\,\mathrm{m}\leq d_{2D}\leq d_{BP}^{\prime}$, the PL is given by (\ref{2}). If $d_{BP}^{\prime}\leq d_{2D}\leq 5\,\mathrm{km}$, the PL is given by (\ref{3}).  In the NLOS case, the PL is determined by comparing the $PL_{UMi-LOS}$ with $PL_{UMi-NLOS}^{\prime}$, and the maximum one gets selected where $PL_{UMi-NLOS}^{\prime}$ is represented by (\ref{5}). Finally, the maximum one from LOS and NLOS, along with SF, is selected as the relevant channel's overall PL. SF is considered as a log-normally distributed random variable with a mean of $0\,dB$ and a standard deviation of $\sigma_{SF}$ , as mentioned in (\ref{shadowfading}). This PL model enables the characterization of wireless channels in UMi environments for system analysis and validation.

\section{Performance metrics}
\label{Performance metrics}
This section presents various performance metrics employed to evaluate the performance of the proposed model. These metrics provide valuable insights into the SNR, throughput, latency, and reliability aspects of our proposal. By rigorously evaluating the performance, we gain a deeper understanding of the capabilities and limitations of our proposed system, enabling us to make informed comparisons with existing systems and highlight the contributions of our research.

\subsection{Signal-to-noise ratio (SNR)}

SNR generically means the dimensionless ratio of the signal power to the noise power, often quantifying the quality of a signal by comparing the power of the signal to the power of the noise present in the system \cite{b36}. For a wireless link, the received SNR can be given by, 

\begin{align}
     SNR^{i,j} &=  \frac{{P_{r}}^{i,j}}{N_{o} \times BW} 
\end{align}
where,
\begin{description}
        \item[$SNR^{i,j}$] \hspace{0.2cm} = Received SNR of the link between $i^{th}$ transmitting vehicle and $j^{th}$ receiving vehicle/SM
        \item[${P_{r}}^{i,j}$] \hspace{0.2cm} = Received power at $j^{th}$ receiving vehicle/SM from $i^{th}$ vehicle
        \item [$BW$] \hspace{0.2cm} = Channel bandwidth
        \item[$N_{o}$] \hspace{0.2cm} = Noise power density
\end{description} 

\subsection{Throughput}
We consider channel capacity proposed by the Shannon-Hartley theorem as the throughput of the channel, which is the maximum data rate that can be transmitted over wireless channels with asymptotically small error probability, assuming no constraints on delay or complexity of the encoder and decoder as represented by (\ref{throughput}) \cite{b37}.

\begin{align}
    C^{i,j} &=  log_{2}(1 + SNR^{i,j}) \label{throughput} 
\end{align}
where,
\begin{description}
    \item[$C^{i,j}$] = Shannon's channel capacity of the link between $i_{th}$ transmitting vehicle and $j_{th}$ receiving vehicle/SM
\end{description}

\subsection{Latency}
End-to-end (E2E) latency refers to the time it takes to transfer a given piece of information from a source to a destination, measured at the application level from the moment it is transmitted by the source to the moment it is received at the destination \cite{b38}. Latency includes the over-the-air propagation delay, data transmission delay at the transmitter, data reception delays at the receiver, and processing/computing delay at different nodes. We define the E2E latency as given below, which is used for performance evaluation.

\begin{align}
    t_{prop} &= \frac{d^{i,j}}{c} \label{propagation}\\
    t_{trans} &= \frac{\beta}{C^{i,j}}\\
    t_{process} &= \begin{cases}2t_{trans} & single \hspace{0.1cm} hop\\2t_{trans} + (N_{H} - 1) \\ \times(t_{transfer} + 2t_{trans}) & multi \hspace{0.1cm} hop\end{cases} \label{processingdelay} \\
    t_{latency} &= t_{prop} + t_{process} \label{latency}
\end{align}
where,

\begin{description}
    \item [$t_{prop}$] = Propagation time from $i^{th}$ transmitting vehicle to the $j^{th}$ receiving vehicle/SM
    \item [$d^{i,j}$] = Distance between $i^{th}$ transmitting vehicle and $j^{th}$ receiving vehicle/SM
    \item [$c$] = Light velocity
    \item [$t_{trans}$] = Data packet transmission time from $i^{th}$ transmitting vehicle to $j^{th}$ receiving vehicle/SM, which is also equal to the data packet reception time at the receiver
    \item [$t_{process}$] \hspace{0.1cm} = Total processing delay
    \item[$ \beta$] = Packet size
    \item[$ N_{H}$] = Number of hop
    \item [$t_{latency}$] \hspace{0.1cm} = E2E latency
    \item [$t_{transfer}$]\hspace{0.15cm} = Time required to transfer a packet through an intermediary vehicle
\end{description}

Propagation delay is determined by dividing the distance between the transmitter and the receiver by the light velocity as given in (\ref{propagation}). For transmission delay, the ratio of the size of the data packet to the transmission rate of the corresponding channel is considered. The total processing delay expressions for the single-hop and multi-hop scenarios differ, as discussed below. In the case of single-hop communication, processing delay is determined by (\ref{processingdelay}), where only the corresponding time needed to transmit the packet from the vehicle and receive it at an SM is considered. For multi-hop communication, the time required to transmit a packet by the first transmitting vehicle and to receive it at the final SM is equal to $2t_{trans}$. However, some extra processing delay is attributed to the intermediate vehicles that act like store-and-forward routers. At first, it receives and stores transmitted packets from another vehicle and then forward those packets to the next suitable vehicle or SM. Again $2t_{trans}$ is required by the store and forward mechanism of the intermediate vehicle with an additional $t_{transfer}$ time to transfer a packet through itself from its input port to the output port. Overall E2E latency is determined by adding the propagation and processing delays together in (\ref{latency}). In our work, latency is a key metric for performance evaluation and will be analyzed broadly in the later segment. 

\subsection{Link reliability}
 Reliability refers to the ability of a wireless system or network to consistently and accurately transmit and receive data without disruptions as defined below \cite{b39}.

\begin{align}
    \eta &=  \frac{l_{total} - l_{unrel}}{l_{total} }  \times 100 \label{reliability}
\end{align}
where, 
\begin{description}
    \item[$l_{total}$] = Number of total wireless links
    \item[$l_{unrel}$] = Number of unreliable wireless links
\end{description}

Here, we determine reliability according to (\ref{reliability}). In our system, the transmitted signal from a vehicle reaches a suitable SM or a vehicle with reduced signal power due to the effect of path loss and fading. If this received power is within the sensitivity limit of the receiver, then successful data reception is possible, and we consider this one as a reliable link; otherwise, the link is treated as an unreliable one. Finally, the percentage of reliable links is considered as the reliability of the system.

\section{Proposed algorithms}
\label{Proposed algorithms}

\begin{figure}[htbp]
\centering
\includegraphics[width=\linewidth]{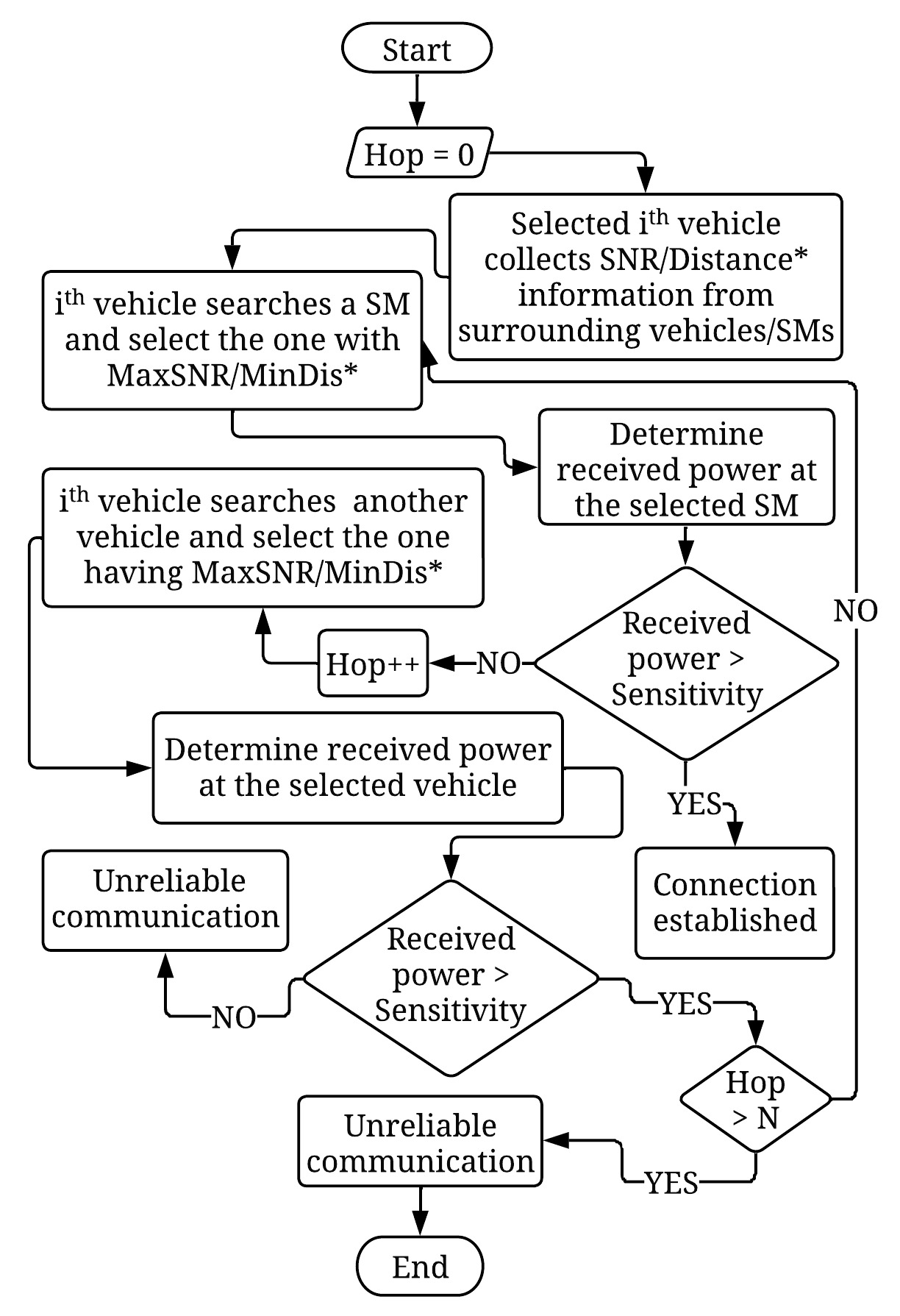}
\caption{Flow diagram of the MaxSNR and MinDis\textsuperscript{*} algorithms.}
\label{Flow diagram of the MAXSNR and $MINDIS^*$ algorithms.}
\end{figure}

We propose two distinct algorithms for vehicle-SM associations: the maximum SNR (MaxSNR) and the minimum distance (MinDis) algorithms. The MaxSNR algorithm maximizes SNR at receiving vehicles/SMs during selection, whereas the MinDis algorithm minimizes the distance between the transmitter and the receiver. Flow diagram of Figure \ref{Flow diagram of the MAXSNR and $MINDIS^*$ algorithms.} provides some insight, and two subsequent subsections comprehensively narrate our proposed algorithms. 

\subsection{MaxSNR algorithm} 

 Let us consider that $i^{th}$ vehicle wants to communicate with an SM to establish a connection with the network. At first, the selected $i^{th}$ vehicle requests all the surrounding  SMs and vehicles to provide information about their individual received SNR from itself, and then through reply messages, it gets the SNR information from the surrounding SMs and vehicles which are in its coverage range. Now, we consider that $i^{th}$ vehicle has received SNR information of $N_{SM}$ SMs and $N_{V}$ vehicles as represented by the following matrices. 
\begin{align}
    \textbf{SNR}_{SM}^{i} &=  [ SNR_{i,1}, SNR_{i,2}, ..., SNR_{i,N_{SM}}]  \label{maxsnrbegin} \\
    \textbf{SNR}_{V}^{i}  &=  [ SNR_{i,1}, SNR_{i,2}, ...,  SNR_{i,N_{V}}] 
\end{align}

At first, the selected $i^{th}$ vehicle searches for MaxSNR from $\textbf{SNR}_{SM}^{i}$ matrix and selects $j^{th}$ SM having maximum received SNR according to (\ref{findingmaxsnr}). In the next step, $i^{th}$ vehicle determines received power at the $j^{th}$ SM ${P_{r}}^{j}$ using (\ref{receivedpowerj}) to check whether it belongs to V2I sensitivity range $P_{V2I}$ or not.

\begin{align}
    SNR^{i,j} &=  max(SNR_{i,1}, SNR_{i,2}, ..... SNR_{i,N_{SM}}) \label{findingmaxsnr} \\
    {P_{r}}^{j} &= SNR^{i,j}\times N_{o}\times BW \label{receivedpowerj}
\end{align}

If ${P_{r}}^{j} > P_{V2I}$, then the vehicle considers it as a reliable communication link and establishes a connection with the selected $j^{th}$ SM. But when ${P_{r}}^{j} < P_{V2I}$, then the $i^{th}$ vehicle considers that there is no suitable SM in the system for reliable communication. Then, instead of searching among SMs, $i^{th}$ vehicle selects the $k^{th}$ vehicle from the $N_{V}$ vehicles providing the highest SNR in matrix $\textbf{SNR}_{V}^{i}$. Then, $i^{th}$ vehicle determines received power at the $k^{th}$ vehicle ${P_{r}}^{k}$ using (\ref{maxsnrend}) to check whether it belongs to V2V sensitivity range $P_{V2V}$ or not.

\begin{align}
    SNR^{i,k} &=  max(SNR_{i,1}, SNR_{i,2}, ..... SNR_{i,N_{V}}) \label{findingmaxsnrvehicle}  \\
    {P_{r}}^{k} &= SNR^{i,k}\times N_{o} \times BW \label{maxsnrend}
\end{align}

If ${P_{r}}^{k} > P_{V2V}$, a link is established between the $i^{th}$ vehicle and the $k^{th}$ vehicle. After that, $k^{th}$ vehicle searches for another suitable SM that receives the maximum SNR from it and repeats through (\ref{maxsnrbegin}) to (\ref{maxsnrend}) up to a specified number of iterations to maximize the possibility of a vehicle finding an appropriate SM. The whole process also repeats for the rest of the vehicles across the network.

\subsection{MinDis algorithm}

The MinDis algorithm works similarly to the MaxSNR algorithm, although MinDis tries to minimize distance instead of maximizing SNR. Here, we consider that a particular vehicle possesses GPS information of the surrounding vehicles and SMs, assuming the same system mentioned in the MaxSNR algorithm. Now $i^{th}$ vehicle can select an appropriate SM for direct connection or an SM using multi-hop communication through other vehicles from the following matrices $\textbf{d}_{SM}^{i}$ and $\textbf{d}_{V}^{i}$. 

\begin{align}
   \textbf{ d}_{SM}^{i} &=  [ d_{i,1}, d_{i,2}, d_{i,3}, ............. d_{i,N_{SM}}]  \label{mindisbegin}\\
  \textbf{  d}_{V}^{i}  &=  [ d_{i,1}, d_{i,2}, d_{i,3},............. d_{i,N_{V}}] 
\end{align}

Here, $\textbf{d}_{SM}^{i}$ and $\textbf{d}_{V}^{i}$ represent distance with all of the $N_{SM}$ SM and $N_{V}$ vehicles from $i^{th}$ vehicle respectively. At first, the selected vehicle searches for the minimum distance from $d_{SM}^{i}$ matrix and select the nearest $j^{th}$ SM from $i^{th}$ vehicle, according to (\ref{findingminimumdis}). Then  $i^{th}$ vehicle determines received power at the $j^{th}$ SM ${P_{r}}^{j}$ using (\ref{receivedpowerjmin}) to check whether it meets the V2I sensitivity limit $P_{V2I}$.

\begin{align}
    d^{i,j} &=  min(d_{i,1}, d_{i,2}, d_{i,3},............. d_{i,N_{SM}}) \label{findingminimumdis} \\
    {P_{r}}^{j} &= P_{t} - PL^{i,j} \label{receivedpowerjmin}
\end{align}
where, \\
$P_{t}$ = Transmit power \\
$PL^{i,j}$ = PL of the link between $i^{th}$ vehicle and $j^{th}$ SM/vehicle\\

If ${P_{r}}^{j} > P_{V2I}$, then the $i^{th}$ vehicle establishes a reliable connection with the $j^{th}$ SM. But when ${P_{r}}^{j} < P_{V2I}$, then the $i^{th}$ vehicle considers it as an unreliable link, and instead of searching among SMs, the $i^{th}$ vehicle selects the nearest $k^{th}$ vehicle from itself as given in (\ref{findingminimumvehicle}). In the next step, $i^{th}$ vehicle determines received power at the $k^{th}$ vehicle ${P_{r}}^{k}$ using (\ref{mindisend}) to check whether it belongs to V2V sensitivity range $P_{V2V}$.

\begin{align}
    d^{i,k} &=  min(d_{i,1}, d_{i,2}, d_{i,3},............. d_{i,N_{V}}) \label{findingminimumvehicle} \\
    {P_{r}}^{k} &= P_{t} - PL^{i,k} \label{mindisend}
\end{align}

If ${P_{r}}^{k} > P_{V2V}$, then a link is established between the $i^{th}$ vehicle and the $k^{th}$ vehicle. After that, $k^{th}$ vehicle searches for another suitable SM that maintains minimum distance with the $k^{th}$ vehicle and repeats through (\ref{mindisbegin}) to (\ref{mindisend}) up to a specified number of iterations. The whole process also repeats for the rest of the vehicles across the network.

\section{Results and discussions}
\label{Results and discussions}

\subsection{Simulation parameters}
\begin{table}
\caption{Communication parameters \cite{b40,b41,b42}}
\label{tab:communication-parameters}
\centering
\begin{adjustbox}{width=\columnwidth}
\begin{tabular}{@{}lcc@{}}
\toprule
\textbf{Item} &
  \begin{tabular}[c]{@{}c@{}}\textbf{Quantity}\\(Proposed SM-based model)\end{tabular} &
  \begin{tabular}[c]{@{}c@{}}\textbf{Quantity}\\(BS-based model)\end{tabular} \\ \midrule
Transmitted Power          & $23\,\mathrm{dBm}$    & $23\,\mathrm{dBm}$     \\
Frequency                  & $5.9\,\mathrm{GHz}$   & $5.9\,\mathrm{GHz}$    \\
Bandwidth                  & $10\,\mathrm{MHz}$    & $10\,\mathrm{MHz}$     \\
Receiver Sensitivity (V2I) & $-92\,\mathrm{dBm}$  & $-103.5\,\mathrm{dBm}$ \\
Receiver Sensitivity (V2V) & $-89\,\mathrm{dBm}$   & $-89\,\mathrm{dBm}$    \\
Packet Size                & $200\,\mathrm{bytes}$ &$200\,\mathrm{bytes}$ \\
Height of BS & $2\,\mathrm{m}$ & $25\,\mathrm{m}$ \\
Height of UT & $1.5\,\mathrm{m}$ & $1.5\,\mathrm{m}$ \\

\bottomrule
\end{tabular}
\end{adjustbox}
\end{table}

Extensive simulations are conducted to evaluate our proposed SM-based V2N architecture. We have developed our own simulation platform using MATLAB. Further simulations are also carried out to compare the performance of our proposed model with that of the conventional BS-based V2N system. Simulation parameters for the proposed SM-based models are selected based on the IEEE 802.11p standard as represented by the $2^{nd}$ column of the Table \ref{tab:communication-parameters}, while for the conventional BS-based method, the LTE-V2X standard is followed as given in the $3^{rd}$ column of the Table \ref{tab:communication-parameters}.

We also consider various physical parameters according to standards to characterize the sub-urban area, along with its road infrastructure and vehicular presence. The default settings of these parameters are presented in Table \ref{tab:physical-parameters}. During simulations, effects arising from the variation of these parameters are analyzed. Latency and reliability stand as the principal evaluation metrics, supplemented by meticulous assessments of throughput and SNR. 

\begin{table}[htbp]
\caption{Urban scenario physical parameters \cite{b43,b44}}
\label{tab:physical-parameters}
\centering
\begin{adjustbox}{width=\columnwidth}
\begin{tabular}{@{}lc@{}}
\toprule
\textbf{Item} & \textbf{Quantity} \\
\midrule
Dimension of the sub-urban area & $2000\,\mathrm{m} \times 2000\,\mathrm{m}$ \\
Plot Size & $200\,\mathrm{m} \times 200\,\mathrm{m}$ \\
Street Width & $20\,\mathrm{m}$ \\
Vehicle Number & $30/\mathrm{road}$ \\
Minimum allowed distance among vehicles & $7\,\mathrm{m}$ \\
SM Density & $2/\mathrm{plot}$ \\
\bottomrule
\end{tabular}
\end{adjustbox}
\end{table}

\subsection{Comparison between the proposed algorithms}

\subsubsection{Vehicle-SM association diagram}

\begin{figure}[htbp]
\centering
\includegraphics[width=\columnwidth]{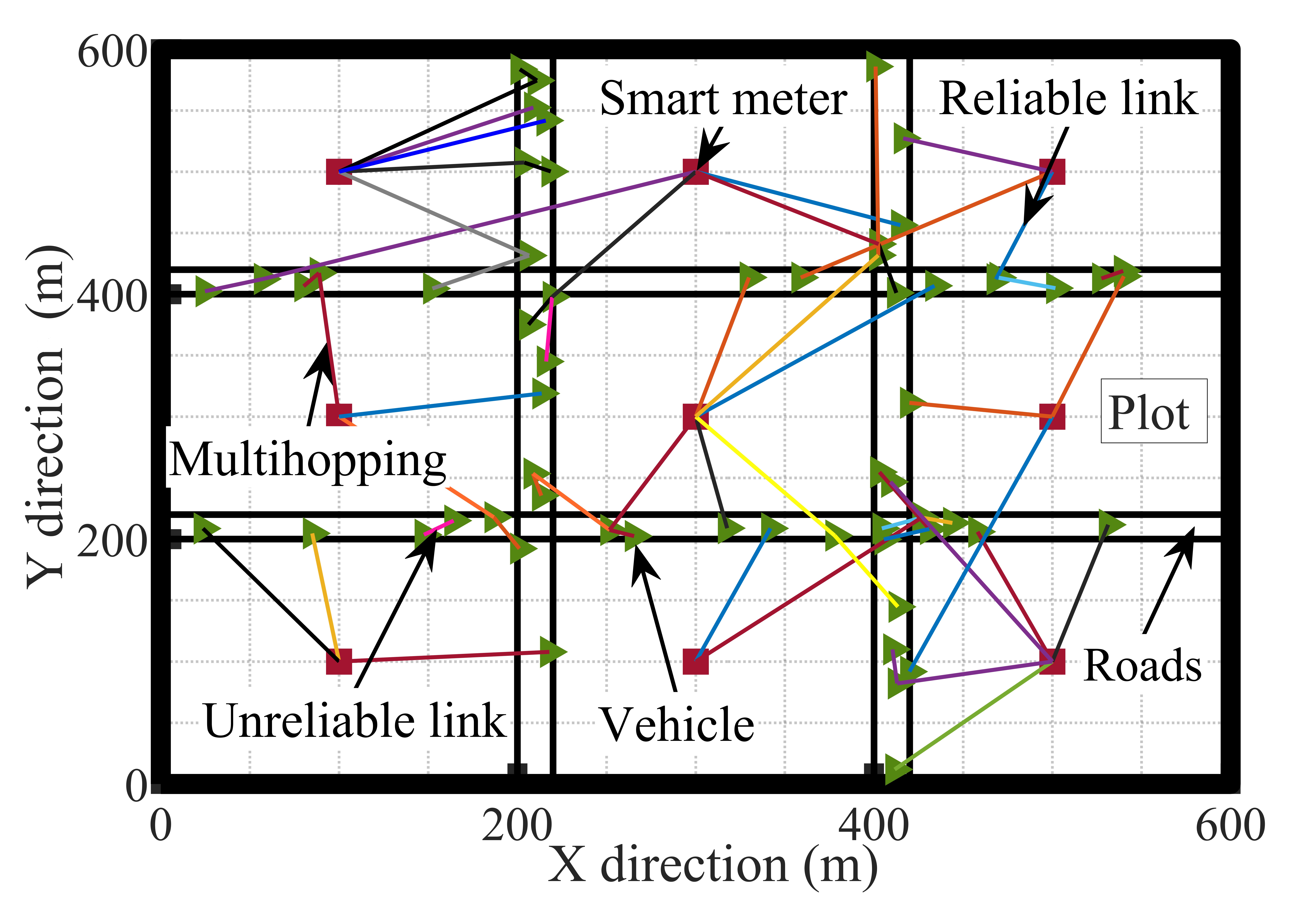}
\caption{Vehicle-SM association under MaxSNR algorithm.}
\label{SM selection diagram according to MAXSNR algorithm.}
\end{figure}

\begin{figure}[htbp]
\centering
\includegraphics[width=\columnwidth]{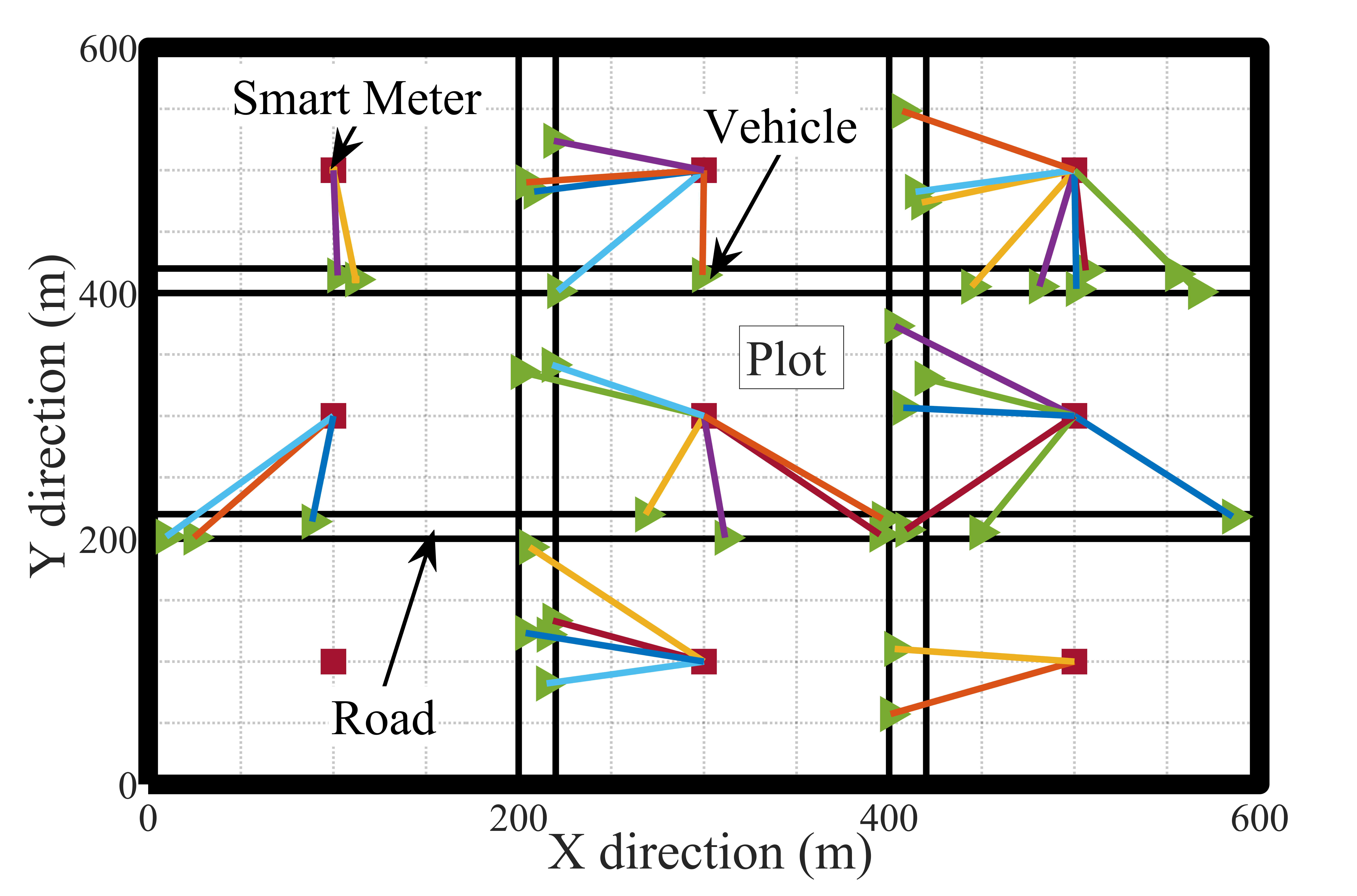}
\caption{Vehicle-SM association under MinDis algorithm.}
\label{SM selection diagram according to MINDIS algorithm.}
\end{figure}

Figure \ref{SM selection diagram according to MAXSNR algorithm.} and Figure \ref{SM selection diagram according to MINDIS algorithm.} below represent the two snapshots of vehicle-SM associations obtained from simulations for both the proposed algorithms, respectively. For the sake of clarity and readability, a reduced number of vehicles and SMs are used for these simulations, while the parameters are used according to Table \ref{tab:communication-parameters} and Table \ref{tab:physical-parameters}. 

In the case of the MinDis algorithm, it is found that vehicles are simply connected with the nearest SM from it. On the other hand, for the MaxSNR algorithm, vehicles are associated with SM on the basis of SNR maximization. Figure \ref{SM selection diagram according to MAXSNR algorithm.} indicates that most vehicles are directly connected with an SM. For some cases, vehicles use other vehicles as a relay to reach up to an SM as the algorithm allows multi-hopping among vehicles for finding a link with better received SNR. The figure also shows very few unreliable links indicating that few vehicles could not find any reliable channel to establish connectivity.

\subsubsection{SNR comparison}
\begin{figure}[htbp]
\centering
\centerline{\includegraphics[width=\columnwidth]{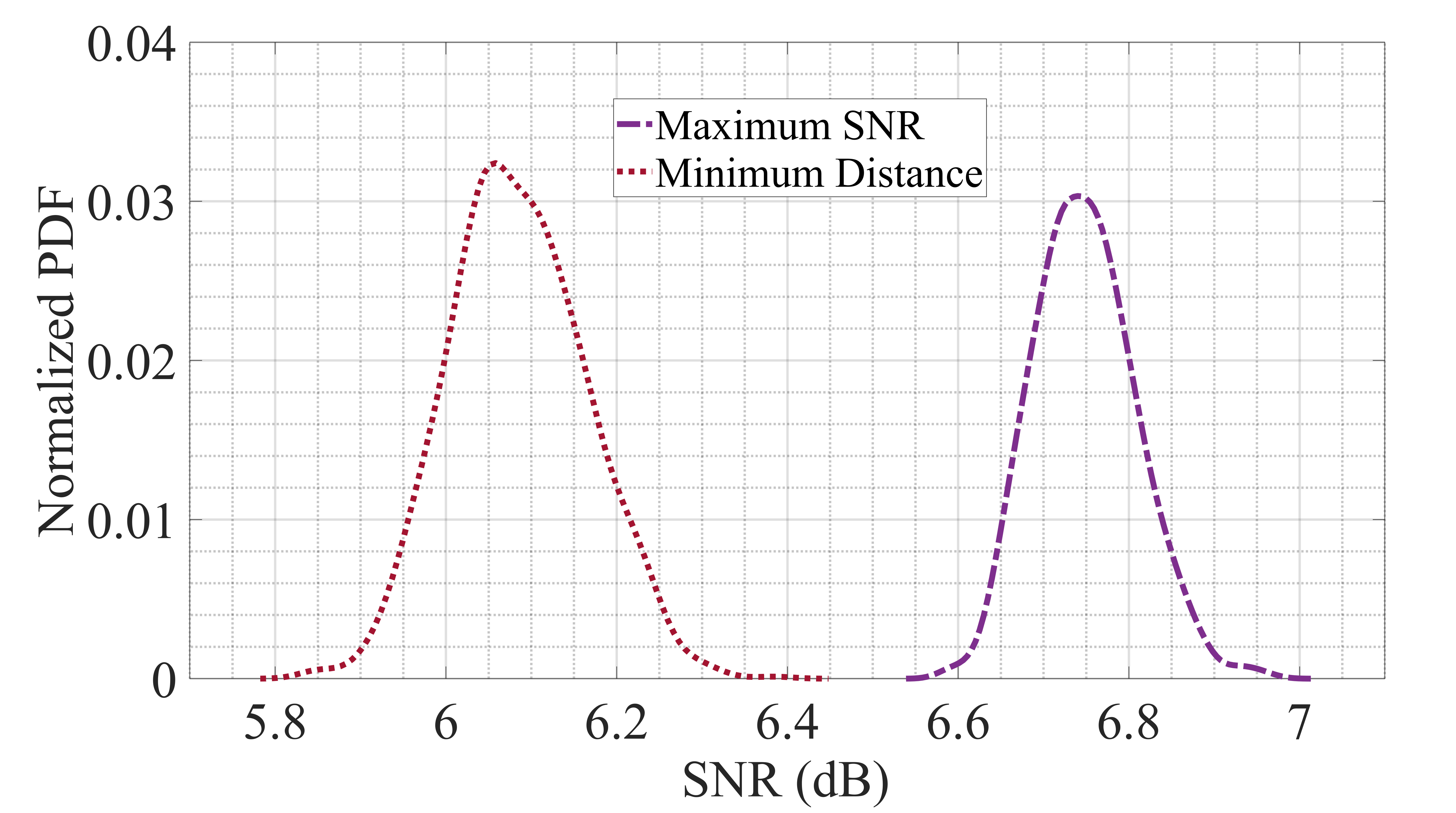}}
\caption{PDF of received SNR at SMs for the proposed algorithms.}
\label{PDF of received SNR at SM for two proposed algorithms.}
\end{figure}

 We evaluate the normalized probability distribution function (PDF) of the received SNR at the SMs under the proposed algorithms, and the results are shown in Figure \ref{PDF of received SNR at SM for two proposed algorithms.}.

 The SNR PDF for the MaxSNR algorithm lies on the right of the MinDis algorithm. More specifically, the majority of values for both algorithms are concentrated around $6.05\,dB$ and $6.75\,dB$, respectively. However, the spread of SNR distribution is notably narrower for the MaxSNR algorithm, with a range of only $0.45\,dB$, compared to the MinDis algorithm's range of $0.65\,dB$. Though the difference isn't that significant, performance in terms of received SNR is better for the MaxSNR algorithm. The superiority of the MaxSNR algorithm lies in the SM selection mechanism, which works based on SNR maximization. Moreover, the lower range and lower standard deviation of received SNRs for the MaxSNR algorithm suggest that the SNR values for the MaxSNR algorithm are tightly clustered around the mean value, increasing the probability of selecting SM with higher SNR for a vehicle, compared to the MinDis algorithm which always selects nearest SM, in terms of SNR, it's pretty random.

\subsubsection{Analysis of multi-hop and single-hop links distribution}

\begin{figure}[htbp]
\centering
\includegraphics[width=\columnwidth]{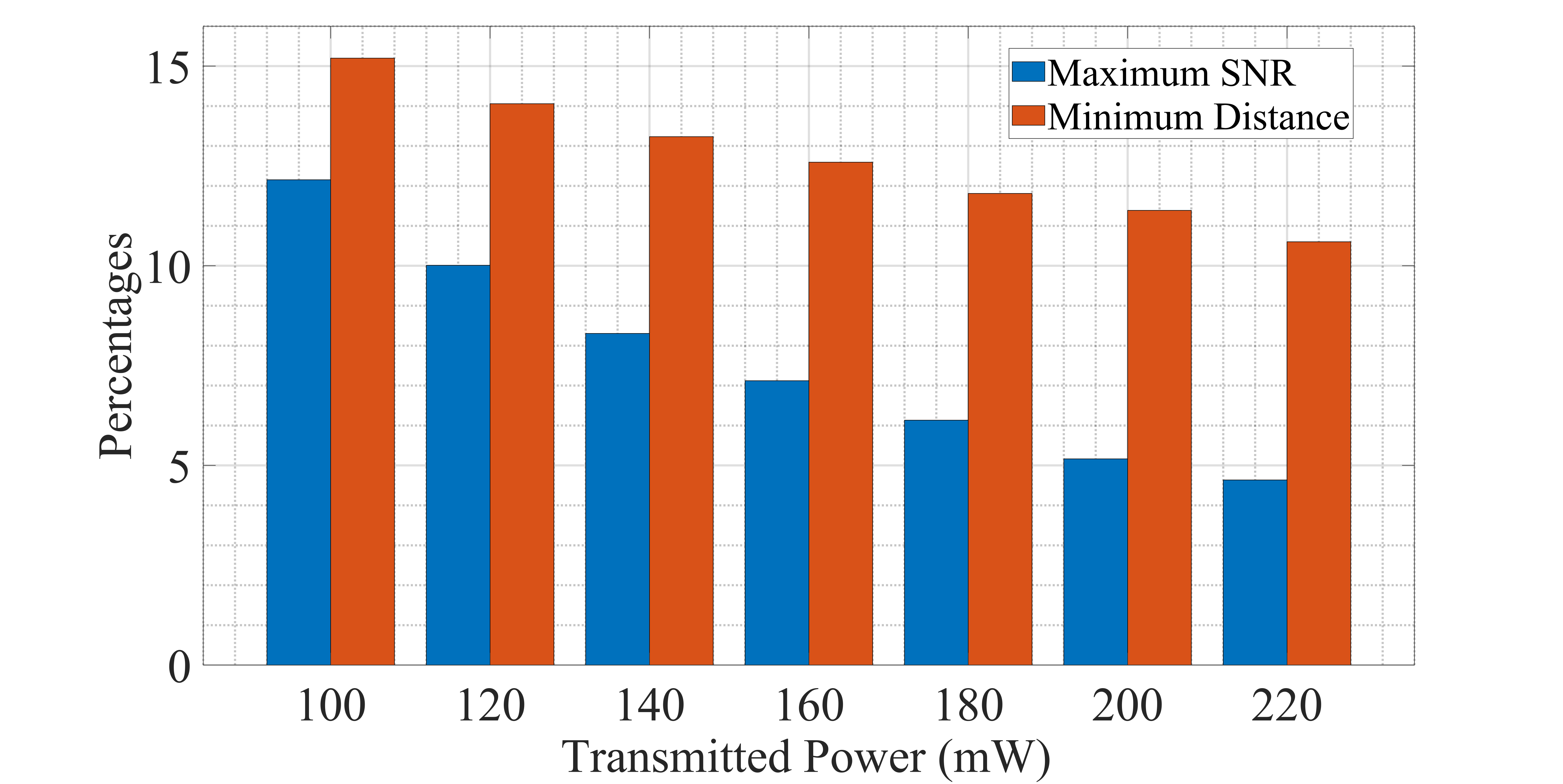}
\caption{Multi-hop link percentages for the two algorithms.}
\label{Multi-hop link percentages for two proposed algorithms.}
\end{figure}

\begin{figure}[htbp]
\centering
\includegraphics[width=\columnwidth]{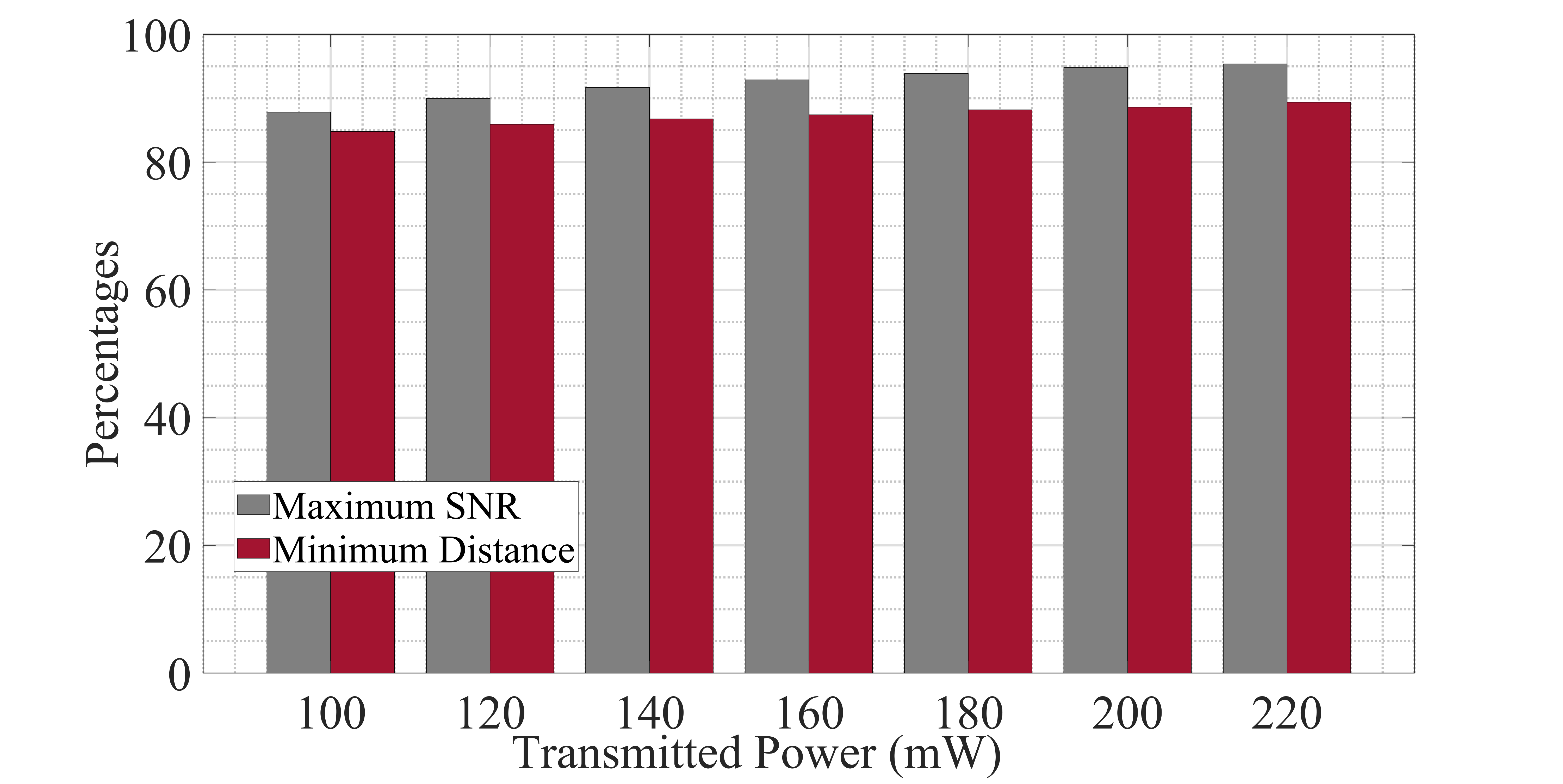}
\caption{Single-hop link percentages for the two algorithms.}
\label{Single-hop link percentages for two proposed algorithms.}
\end{figure}

Our proposed algorithms allow multi-hopping among vehicles in cases where the transmitting vehicle cannot find a suitable SM that maintains received power within the sensitivity range. We also evaluate the percentages of vehicles that directly connect with SMs and those that connect via other vehicles. Corresponding outcomes are shown in Figure \ref{Multi-hop link percentages for two proposed algorithms.} and Figure \ref{Single-hop link percentages for two proposed algorithms.}.

For the MaxSNR algorithm, the percentage of multi-hop links is consistently lower compared to that under the MinDis algorithm. In the case of single-hop links, the scenario is the opposite. As the MaxSNR algorithm ensures better received SNR at SMs, as a consequence, it also increases the probability of directly finding an SM for a vehicle. Additionally, a trade-off is observed with respect to transmit power. As transmit power increases, the probability of directly finding an SM from a particular vehicle also increases due to the expanded coverage resulting from the increased transmit power. Consequently, there is a decrease in the percentage of multi-hop links and an increase in single-hop links, as notified in the plots. A lower overall multi-hop link percentage also indicates a reduced power consumption for signal transmission due to fewer instances of the relaying of signal transmission.

\subsubsection{Reliability and latency comparison}

\begin{figure}[htbp]
\centering
\includegraphics[width= \columnwidth]{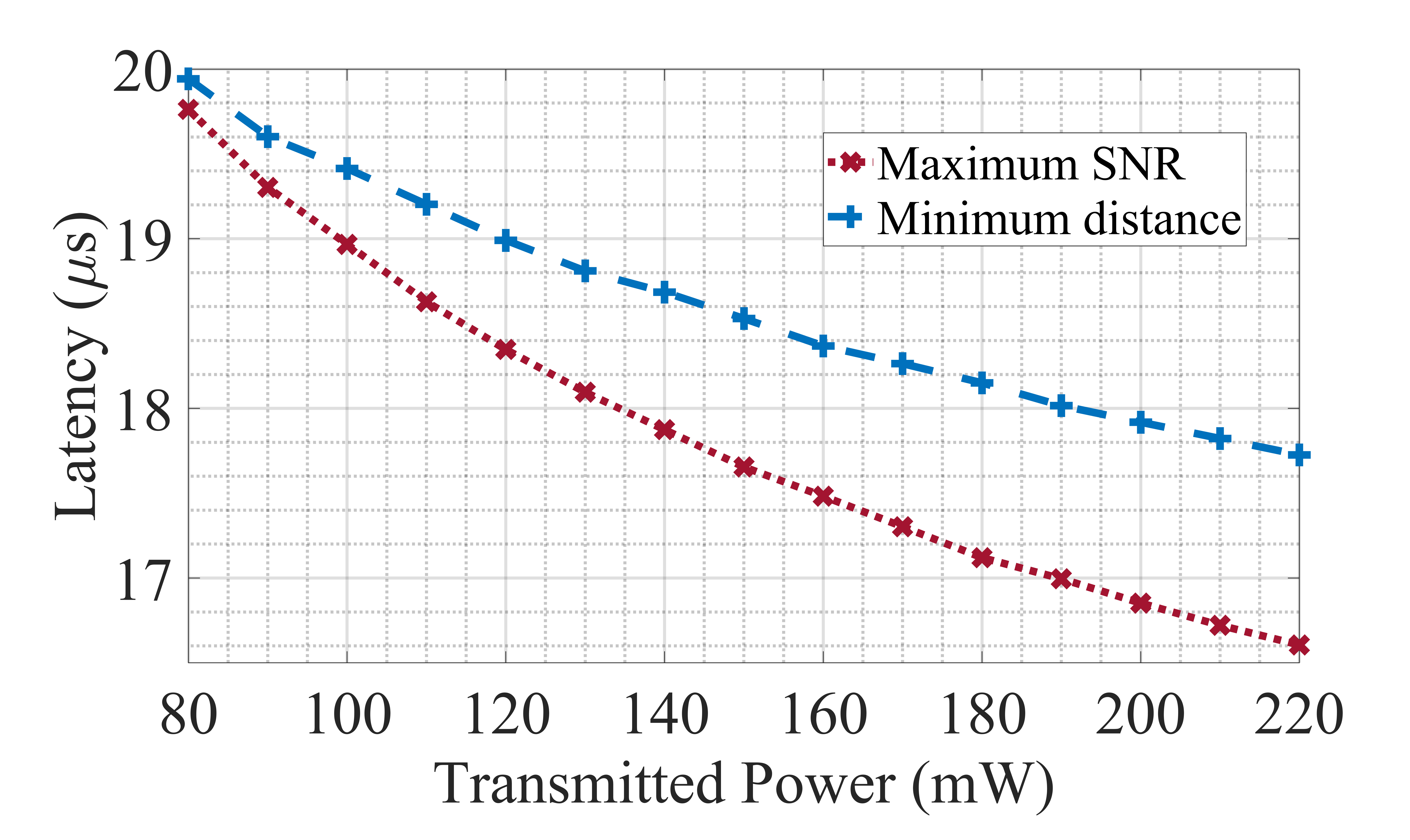}
\caption{Latency comparison for the proposed algorithms.}
\label{Latency comparison for two proposed algorithms.}
\end{figure}
\begin{figure}
\centering
\includegraphics[width= \columnwidth]{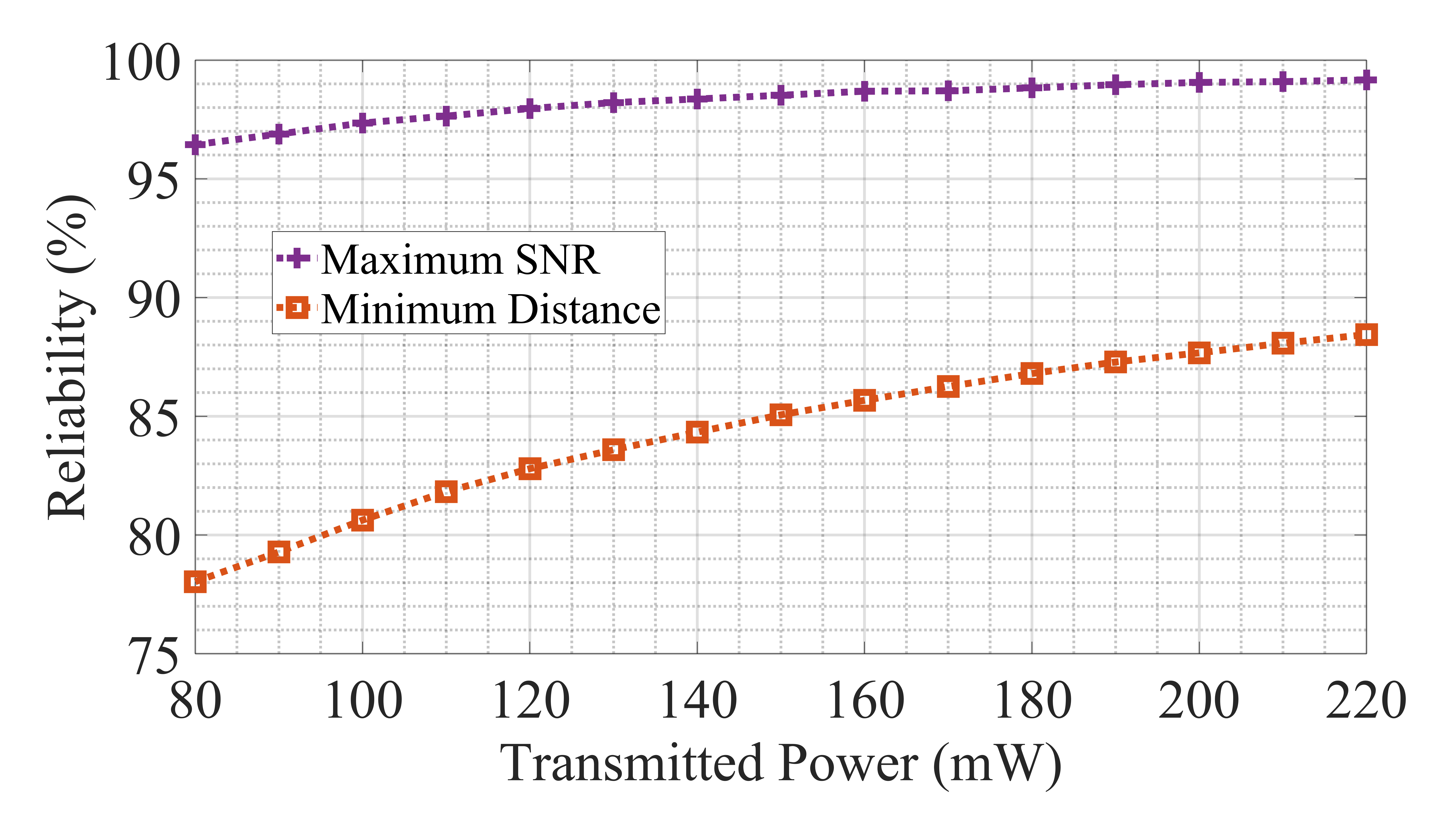}
\caption{Reliability comparison for two proposed algorithms.}
\label{Reliability comparison for two proposed algorithms.}
\end{figure}
Latency performance of the proposed algorithms is illustrated in Figure \ref{Latency comparison for two proposed algorithms.}. Latency under the MaxSNR algorithm consistently remains below that of the MinDis algorithm. However, it is notable that the difference in latency between the two algorithms is relatively small, ranging between $0.1 \,\mu s$ to $1.5 \,\mu s$, as shown in the figure. The MinDis Algorithm selects SMs or vehicles based on the distance among themselves, resulting in a comparatively low propagation delay attribute of latency. On the other hand, the MaxSNR algorithm selects vehicles or SM based on the SNR of the link, resulting in a comparatively low transmission delay attribute. These two phenomena tend to neutralize each other's differences, resulting in a small difference.

Comparison in terms of reliability further consolidates the superiority of the MaxSNR algorithm as evident in Figure \ref{Reliability comparison for two proposed algorithms.}. Reliability for the MaxSNR algorithm is found to be significantly higher compared to that of the MinDis algorithm. At an operating transmit power of $200\,\mathrm{mW}$, reliability is approximately $12.5\%$ higher for the MaxSNR algorithm, as seen in the figure. This difference is reasonable because the MinDis algorithm prioritizes distance over SNR during communications, which leads to a substantial number of unreliable communications due to lower received SNR at SMs. These findings emphasize the effectiveness of the MaxSNR algorithm. Therefore, from now on, we will analyze the results for the MaxSNR algorithm only.

\subsection{Comparison between SM and BS-based system models}

\begin{figure}[htbp]
\centering
\includegraphics[width= \columnwidth]{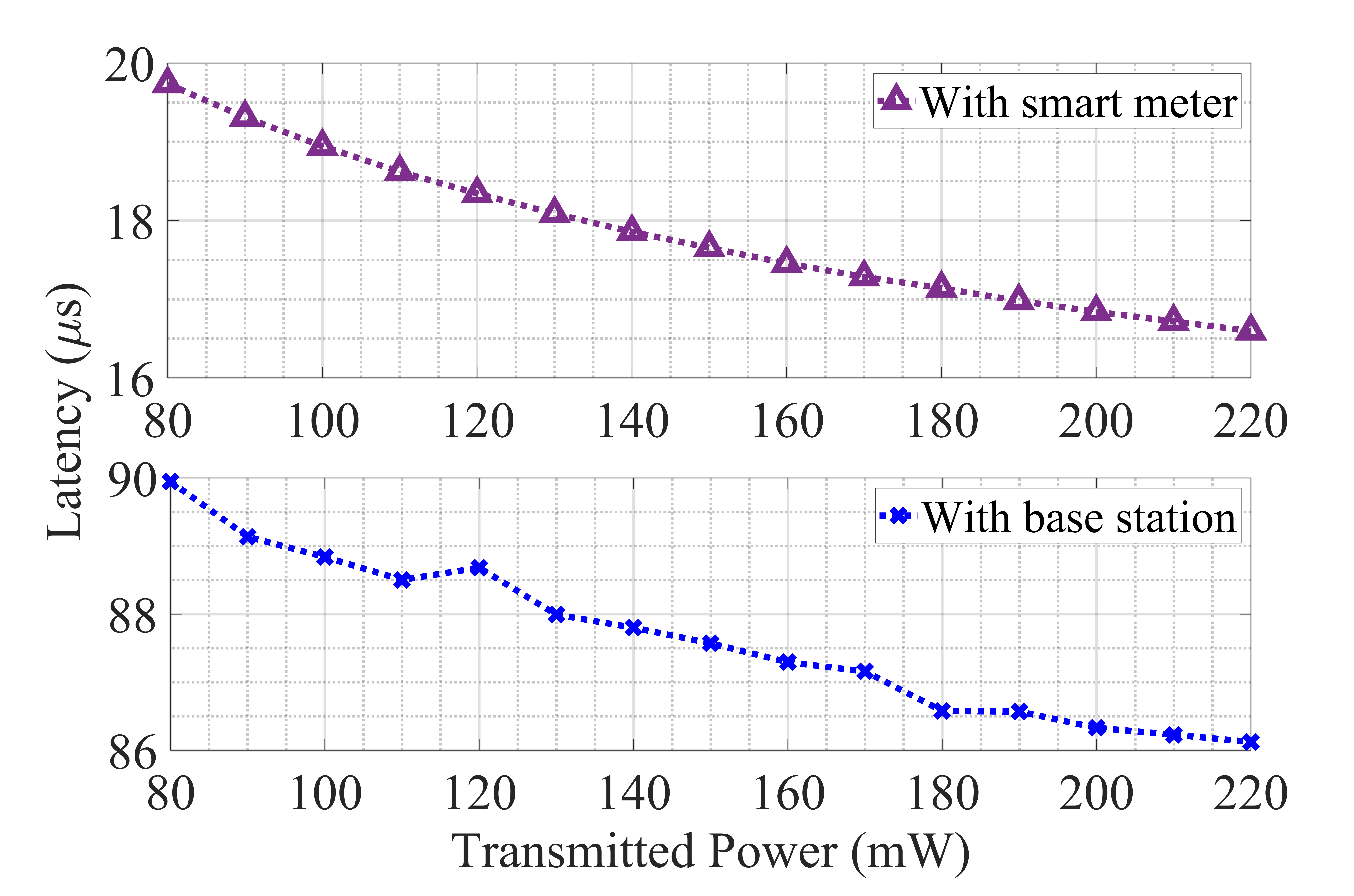}
\caption{Latency comparison between SM and BS-based system models.}
\label{Latency comparison between SM and BS-based communication model.}
\end{figure}

\begin{figure}[htbp]
\centering
\includegraphics[width= \columnwidth]{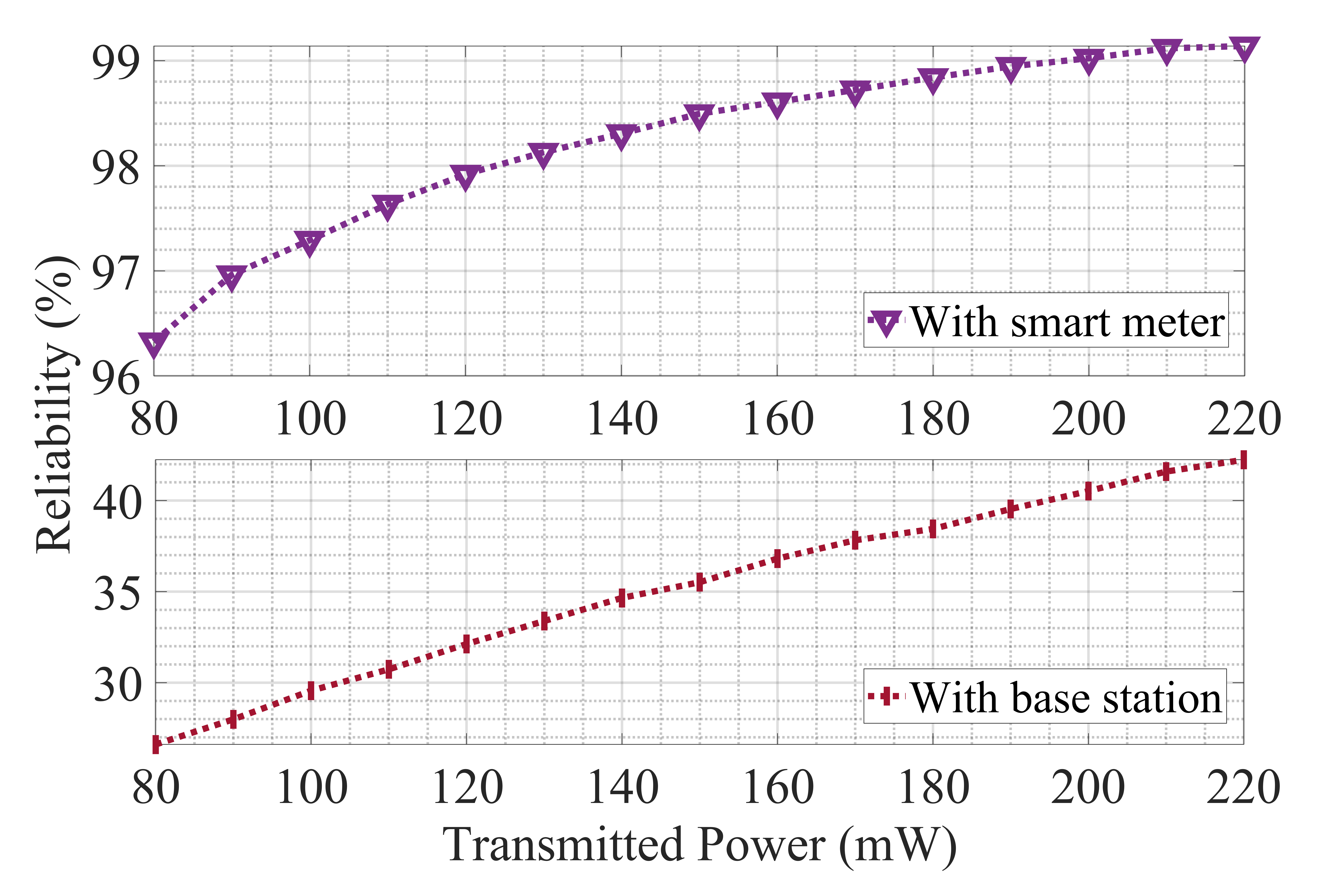}
\caption{Reliability comparison between SM and BS-based system model.}
\label{Reliability comparison between SM and BS-based communication model.}
\end{figure}

We aim to propose a new network architecture utilizing SMs as distributed APs of network connectivity, which would be an alternative to the conventional BS-based model to enhance reliability and latency. For comparison, we utilize the parameters from Table \ref{tab:communication-parameters} and Table \ref{tab:physical-parameters} for simulations. Additionally, for the BS-based model, we consider two BSs uniformly placed in the middle of the first half ($1000\,m\times1000\,m$) and the second half ($1000\,m\times1000\,m$) of the suburban area. Figure \ref{Latency comparison between SM and BS-based communication model.} and Figure \ref{Reliability comparison between SM and BS-based communication model.} represent the latency and reliability comparisons, respectively.

For the proposed SM-based system, latency drops to around $16.8\,\mu s$ when operating at a transmit power of $200\,mW$, whereas, for the BS-based model, the latency is approximately $86.4\,\mu s$. By incorporating SMs into the system instead of BS, we also observe a substantial improvement in reliability. The reliability increases from approximately $40\%$ in the BS-based model to $99\%$ in the SM-based model at operating transmit power of $200\,mW$. This improvement in reliability and latency is attributed to the inclusion of SMs, as they provide distributed APs at a closer distance for connecting vehicles to the network.

\subsection{Effect of plot size}

\begin{figure}[htbp]
\centering
\includegraphics[width= \columnwidth]{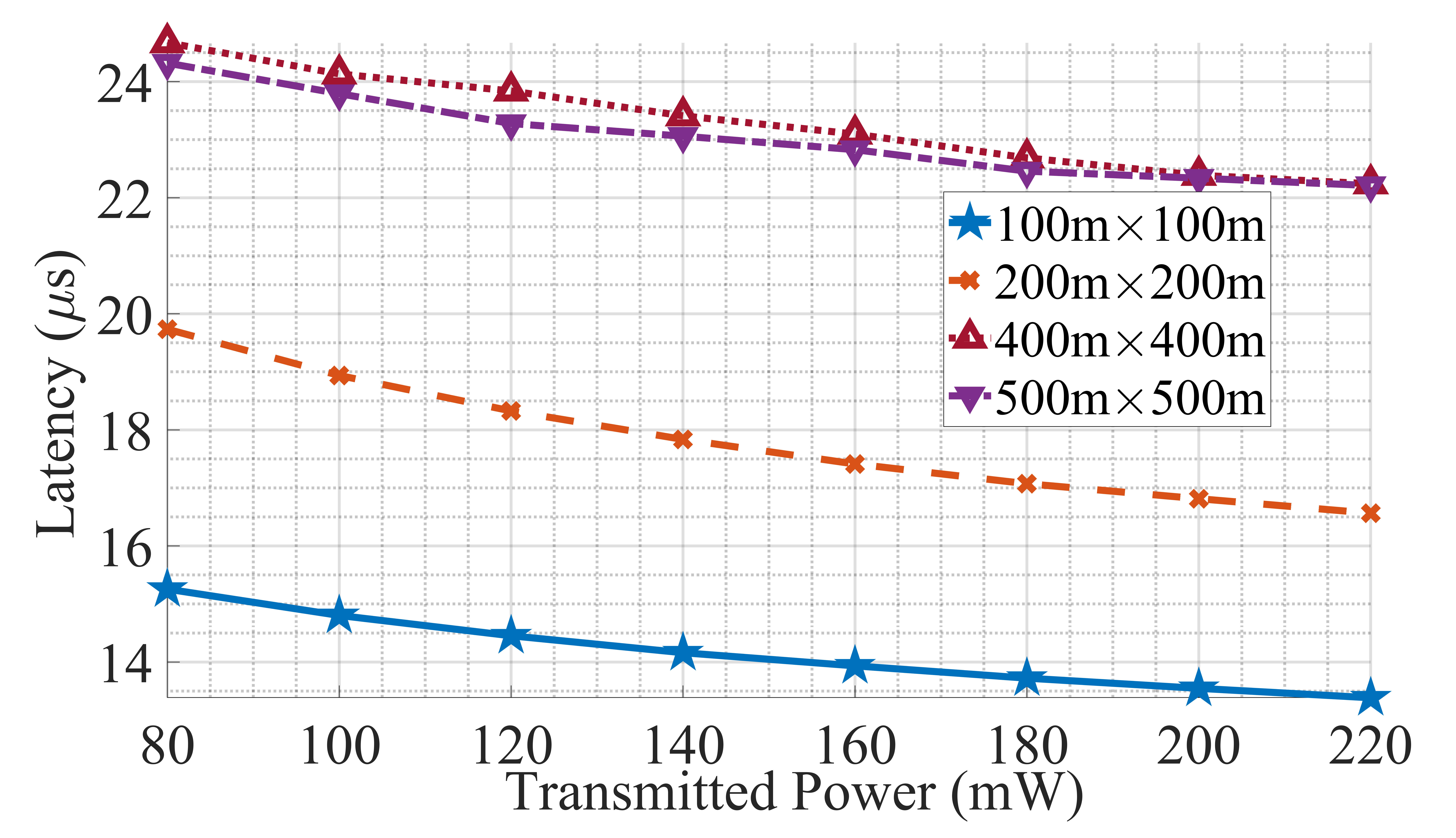}
\caption{Effect of plot size on latency.}
\label{Effect of plot size on latency.}
\end{figure}

\begin{figure}[htbp]
\centering
\includegraphics[width= \columnwidth]{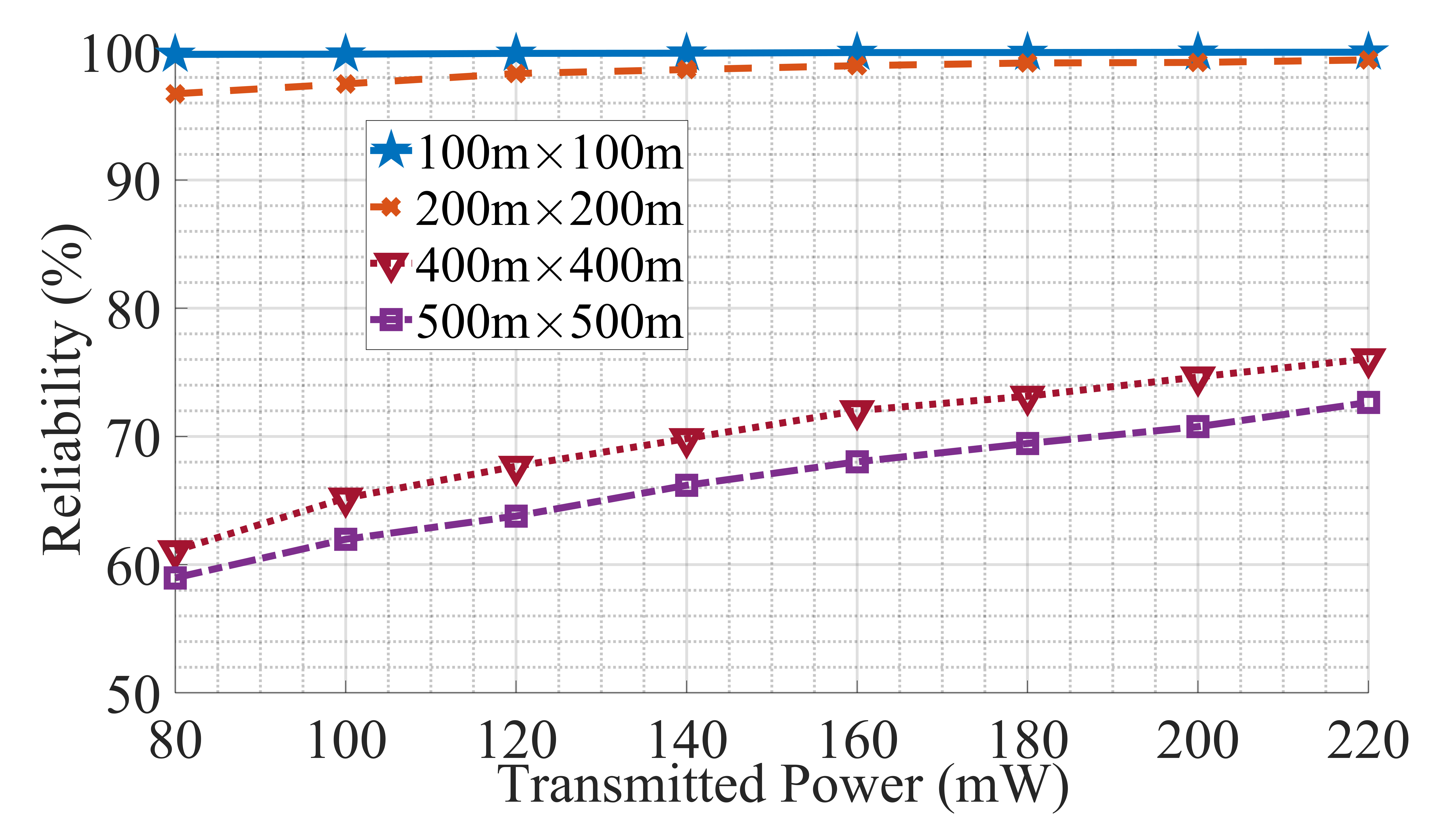}
\caption{Effect of plot size on reliability.}
\label{Effect of plot size on reliability.}
\end{figure}

We conduct simulations to investigate the impact of plot size variations in the sub-urban area on reliability and latency while keeping all other parameters as mentioned in Table \ref{tab:communication-parameters} and Table \ref{tab:physical-parameters}. The purpose is to analyze how changes in plot size affect the QoS metrics. The findings are presented in Figure \ref{Effect of plot size on latency.} and Figure \ref{Effect of plot size on reliability.}.

The results indicate that larger plot sizes have a negative impact on latency and reliability. This observation can be attributed to the fact that while the plot size increases, the number of SMs, which is an essential system parameter, remains the same. Consequently, as the plot size increases, distances between vehicles and SMs keep increasing, leading to reduced SNR. This results in an increase in latency and a decrease in reliability as larger plot sizes pose challenges in maintaining a stable and efficient connection. These findings highlight the significance of considering plot size as a crucial factor when designing and implementing proposed SM-based networks in suburban areas.

\subsection{Effect of vehicle density}

Vehicle congestion is a prevalent issue in densely populated cities, and its impact on latency and reliability is thoroughly analyzed. We examine the effect
\begin{figure}[htbp]
\centering
\includegraphics[width= \columnwidth]{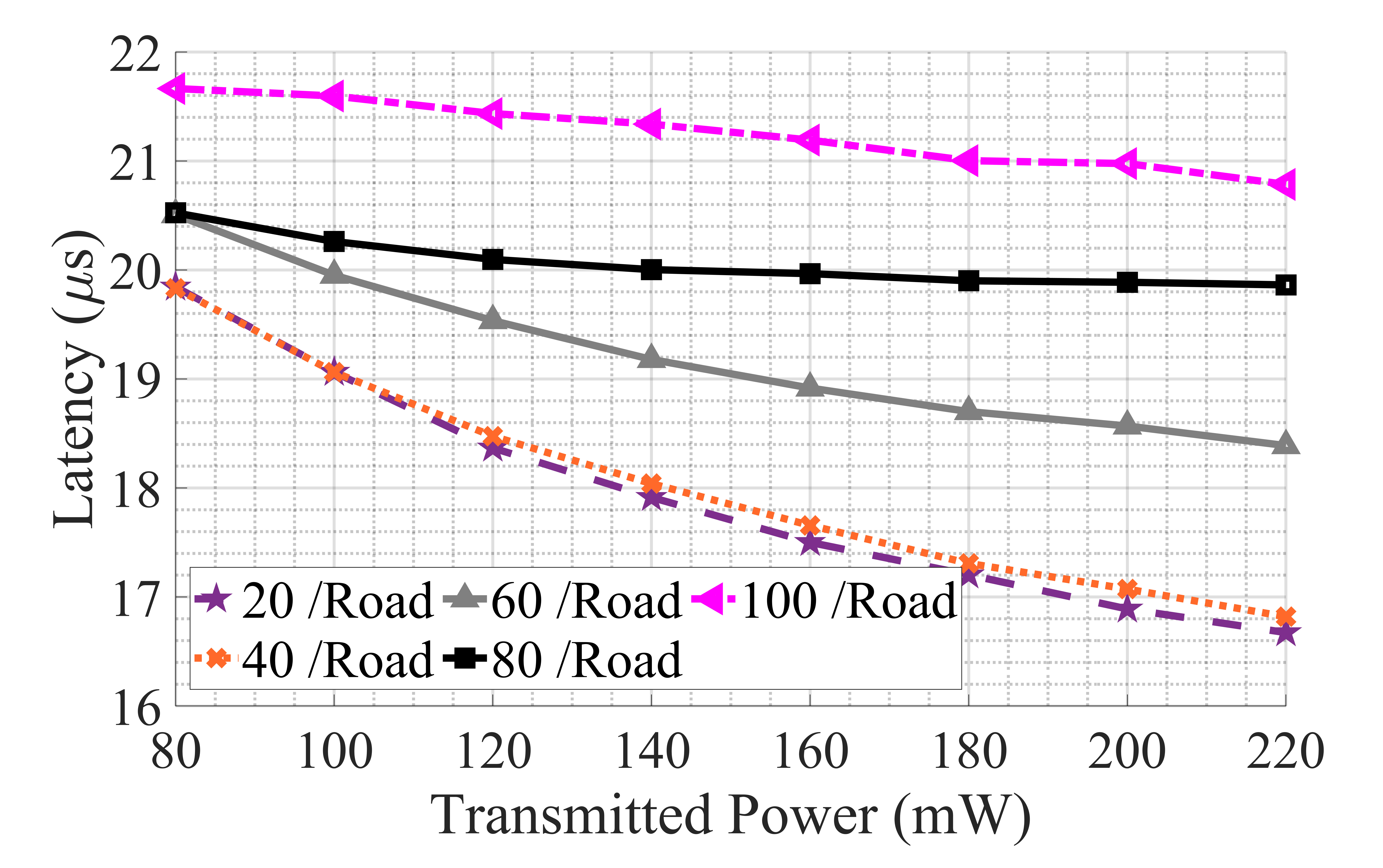}
\caption{Effect of vehicle density on latency.}
\label{Effect of vehicle density on latency.}
\end{figure}

by varying vehicle density per road, ranging from $20$ to $100$ vehicles (each road is $2000\,m$ long), encompassing a wide range of low to high-density scenarios while maintaining other parameters as it was previously.

\begin{figure}[htbp]
\centering
\includegraphics[width= \columnwidth]{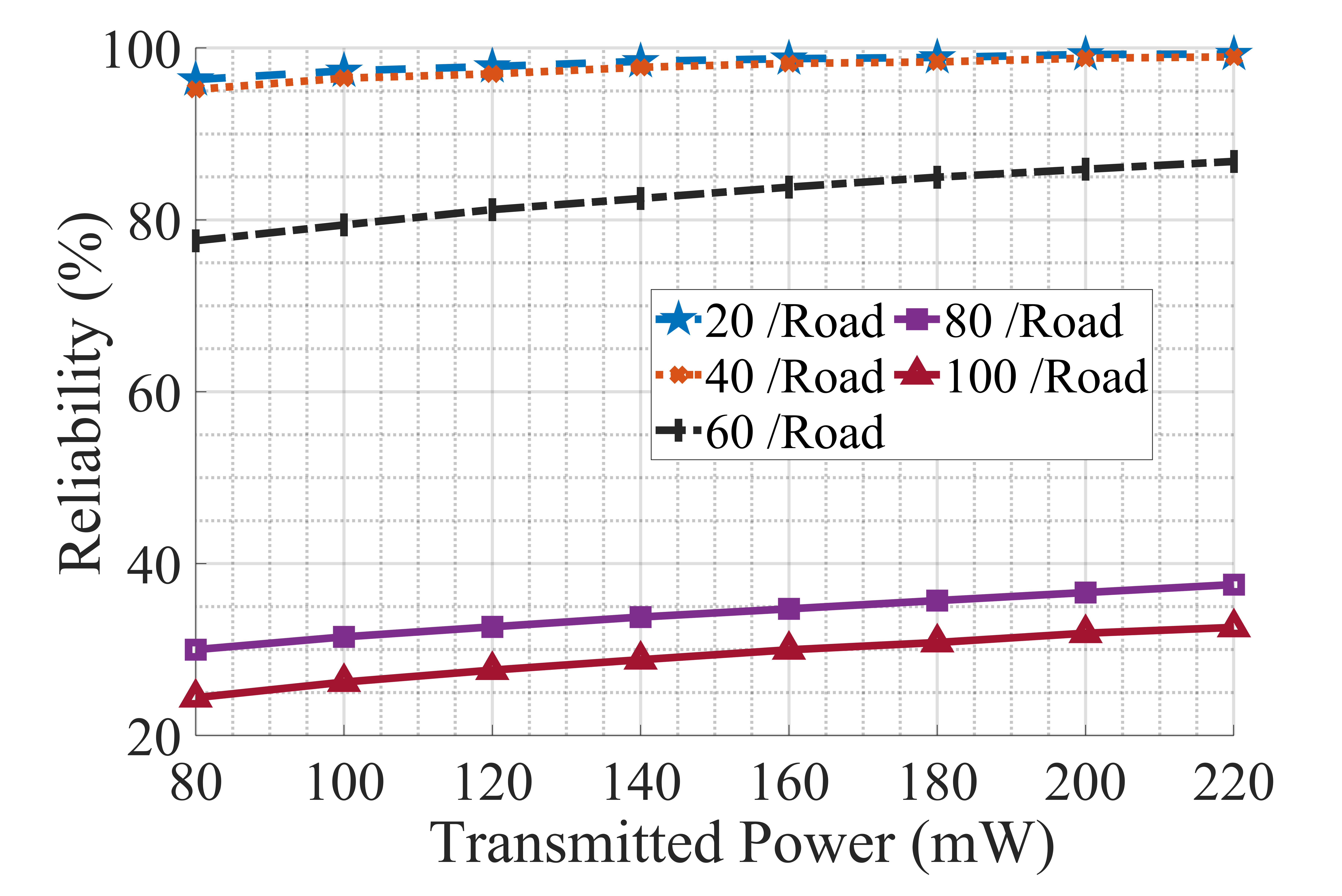}
\caption{Effect of vehicle density on reliability.}
\label{Effect of vehicle density on reliability.}
\end{figure}

Throughout our simulations, we maintain a constant number of SMs per plot, implying that an increased number of vehicles had to contend for the same amount of resources to establish connectivity with the network. This resulted in a degradation of system performance, characterized by increased latency and reduced reliability as demonstrated in Figure \ref{Effect of vehicle density on latency.} and Figure \ref{Effect of vehicle density on reliability.}. Under typical vehicle densities of $20$ or $40$ vehicles per road, the system performed within desired standards. However, as vehicle congestion intensified, reliability plummeted to as low as $30\%$. However, the impact on latency is not that much as the connections are established only for those links offering SNR within the sensitivity range. An increment of $3\,\mu s$ or $4\,\mu s$ in latency is notified when the vehicle density increases from $20$ or $40$ vehicles per road to $80$ or $100$ vehicles per road.

\subsection{Effect of SM density}

\begin{figure}[htbp]
\centering
\includegraphics[width=\columnwidth]{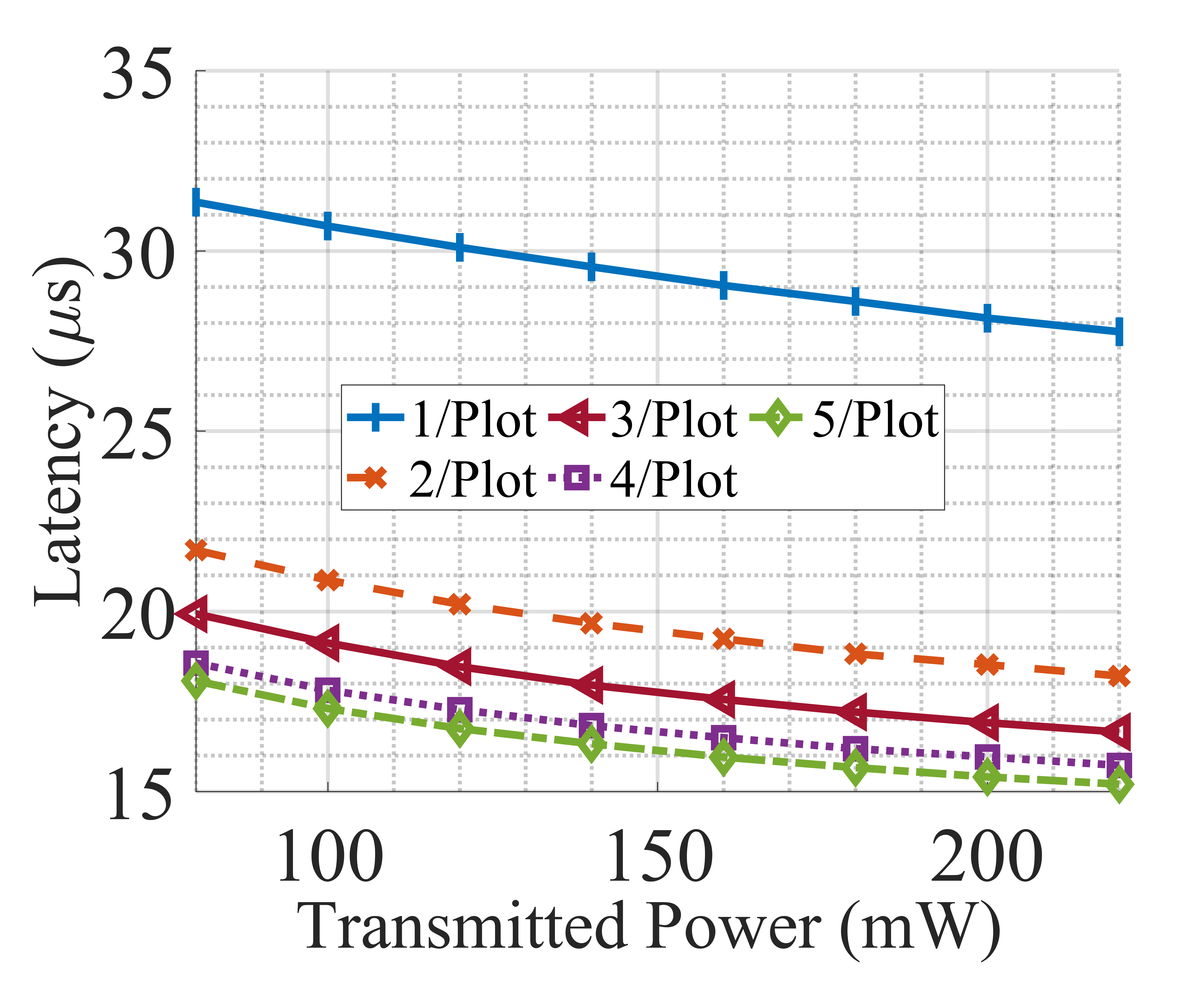}
\caption{Effect of SM density on latency.}
\label{Effect of SM density on latency.}
\end{figure}

\begin{figure}[htbp]
\centering
\includegraphics[width=\columnwidth]{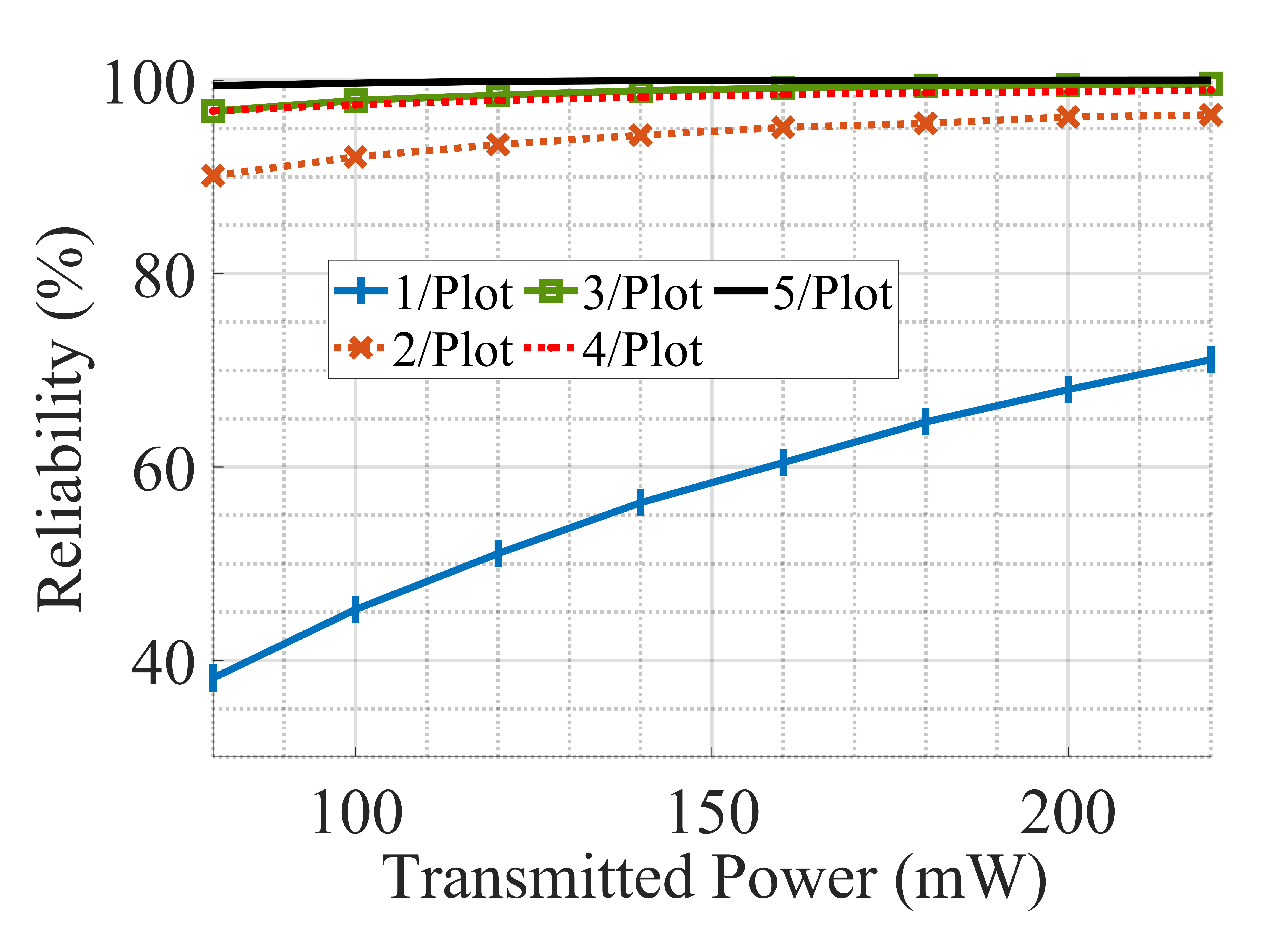}
\caption{Effect of SM density on reliability.}
\label{Effect of SM density on reliability.}
\end{figure}

The number of SMs per plot or SM density has a significant impact on the reliability and latency of the system. To ensure a homogeneous communication environment, the SMs are uniformly distributed within each $200\,m\times200\,m$ plot, and we observe that variations in their density lead to different outcomes as shown in Figure \ref{Effect of SM density on latency.} and Figure \ref{Effect of SM density on reliability.}.

Three, four, and five SMs per plot achieve the best possible QoS metrics, including reduced latency and increased reliability. Remarkably, deploying five SMs per plot achieves nearly $100\%$  reliability even at low transmit power levels. Furthermore, our analysis demonstrates a clear trade-off associated with SM density. Increasing the number of SMs per plot positively impacts the communication experience, enhancing reliability and reducing latency. This observation aligns with the reasoning that a higher number of houses per plot corresponds to a greater abundance of SMs, thus providing additional network resources for vehicles. Consequently, a denser distribution of SMs facilitates an improved communication experience.

\section{Conclusions}
\label{Conclusions}

This paper has presented an SM-based CN architecture of V2N communications for URLLC. Two distinct algorithms - MaxSNR and MinDis, with the option of both single-hop and multi-hop communications, have been proposed for associating the vehicles with the SMs. Extensive simulations have been carried out to evaluate the performance of the proposed architecture and algorithms in terms of various performance metrics, including latency and reliability. It has been found that the MaxSNR algorithm has offered superior results in terms of all performance metrics. Additionally, we have compared our proposed architecture with the conventional BS-based system model. Our proposed architecture has clearly outperformed the conventional one in terms of reliability and latency. Finally, we have presented an in-depth analysis of the impact of vehicle density, SM density, and plot size on the latency and reliability performance of the proposed architecture. In our future works, we will focus on the optimization of the proposed architecture and the development of new algorithms considering QoS constraints, limited radio resources, and vehicular dynamics.


\begin{thebibliography}{00}

\bibitem{b1} A. Rejeb, K. Rejeb, S. Simske, H. Treiblmaier, and S. Zailani, “The big picture on the internet of things and the smart city: a review of what we know and what we need to know,” Internet of Things, vol. 19, p. 100565, Aug. 2022, doi: https://doi.org/10.1016/j.iot.2022.100565.
‌

\bibitem{b2} “Smart cities ranking: an effective instrument for the positioning of the cities?,” ACE: Architecture, City and Environment, Feb. 2010, doi: https://doi.org/10.5821/ace.v4i12.2483.
‌

\bibitem{b3} Amendment of Parts 2 and 90 of the Commission’s Rules to Allocate the 5.850-5.925 GHz Band to the Mobile Service for Dedicated Short Range Communications of Intelligent Transportation Services, ET Docket No. 98-95, Report and Order, 14 FCC Rcd 18221 (1999).

\bibitem{b4} S. Eichler, "Performance Evaluation of the IEEE 802.11p WAVE Communication Standard," 2007 IEEE 66th Vehicular Technology Conference, Baltimore, MD, USA, 2007, pp. 2199-2203, doi: 10.1109/VETECF.2007.461.

\bibitem{b5} S. Hakak et al., “Autonomous vehicles in 5G and beyond: A survey,” Vehicular Communications, vol. 39, p. 100551, Feb. 2023, doi: https://doi.org/10.1016/j.vehcom.2022.100551.
‌
‌
\bibitem{b6} C. She et al., "A Tutorial on Ultrareliable and Low-Latency Communications in 6G: Integrating Domain Knowledge Into Deep Learning," in Proceedings of the IEEE, vol. 109, no. 3, pp. 204-246, March 2021, doi: 10.1109/JPROC.2021.3053601.

\bibitem{b7} Z. Li, M. A. Uusitalo, H. Shariatmadari and B. Singh, "5G URLLC: Design Challenges and System Concepts," 2018 15th International Symposium on Wireless Communication Systems (ISWCS), Lisbon, Portugal, 2018, pp. 1-6, doi: 10.1109/ISWCS.2018.8491078.

\bibitem{b8} G. R. Barai, S. Krishnan and B. Venkatesh, "Smart metering and functionalities of smart meters in smart grid - a review," 2015 IEEE Electrical Power and Energy Conference (EPEC), London, ON, Canada, 2015, pp. 138-145, doi: 10.1109/EPEC.2015.7379940.

\bibitem{b9} K. C. Chen, Y. Peng, N. R. Prasad, Y. C. Liang, and S. Sun, “Cognitive radio network architecture,” Jan. 2008, doi: https://doi.org/10.1145/1352793.1352817.

\bibitem{b10} S. Haykin, "Cognitive radio: brain-empowered wireless communications," in IEEE Journal on Selected Areas in Communications, vol. 23, no. 2, pp. 201-220, Feb. 2005, doi: 10.1109/JSAC.2004.839380.

‌\bibitem{b11} E. Biglieri, Principles of Cognitive Radio. Cambridge University Press, 2013. Accessed: Jul. 30, 2023. [Online].

\bibitem{b12} Y. Kabalci, “A survey on smart metering and smart grid communication,” Renewable and Sustainable Energy Reviews, vol. 57, pp. 302–318, May 2016, doi: https://doi.org/10.1016/j.rser.2015.12.114.
‌
\bibitem{b13} "IEEE Guide for Smart Grid Interoperability of Energy Technology and Information Technology Operation with the Electric Power System (EPS), End-Use Applications, and Loads," in IEEE Std 2030-2011 , vol., no., pp.1-126, 10 Sept. 2011, doi: 10.1109/IEEESTD.2011.6018239.

\bibitem{b14} S. Mohagheghi, J. Stoupis, Z. Wang, Z. Li and H. Kazemzadeh, "Demand Response Architecture: Integration into the Distribution Management System," 2010 First IEEE International Conference on Smart Grid Communications, Gaithersburg, MD, USA, 2010, pp. 501-506, doi: 10.1109/SMARTGRID.2010.5622094.

\bibitem{b15} Z. Li, F. Yang and D. Ishchenko, "The standardization of distribution grid communication networks," 2012 IEEE Power and Energy Society General Meeting, San Diego, CA, USA, 2012, pp. 1-8, doi: 10.1109/PESGM.2012.6345482.

\bibitem{b16} “IEC 61968-100 INTERNATIONAL STANDARD NORME INTERNATIONALE Application integration at electric utilities -System interfaces for distribution management - Part 100: Implementation profiles Intégration d’applications pour les services électriques -Interfaces système pour la gestion de distribution - Partie 100: Profils de mise en oeuvre,” 2013. Accessed: Jul. 30, 2023.

\bibitem{b17} R. H. Khan and J. Y. Khan, “A comprehensive review of the application characteristics and traffic requirements of a smart grid communications network,” Computer Networks, vol. 57, no. 3, pp. 825–845, Feb. 2013, doi: https://doi.org/10.1016/j.comnet.2012.11.002.

\bibitem{b18} N. G. Paterakis, O. Erdinç, and J. P. S. Catalão, “An overview of Demand Response: Key-elements and international experience,” Renewable and Sustainable Energy Reviews, vol. 69, pp. 871–891, Mar. 2017, doi: https://doi.org/10.1016/j.rser.2016.11.167.

\bibitem{b19} M. A. Piette et al., “Open Automated Demand Response Communications Specification (Version 1.0),” www.osti.gov, Feb. 28, 2009. https://www.osti.gov/biblio/951952 (accessed Jul. 30, 2023).

\bibitem{b20} S. Mumtaz, K. M. Saidul Huq, M. I. Ashraf, J. Rodriguez, V. Monteiro and C. Politis, "Cognitive vehicular communication for 5G," in IEEE Communications Magazine, vol. 53, no. 7, pp. 109-117, July 2015, doi: 10.1109/MCOM.2015.7158273.

\bibitem{b21} Anna Maria Vegni and D. P. Agrawal, Cognitive Vehicular Networks. CRC Press, 2018.
‌

\bibitem{b22} P. Kolodzy, 2001, October. Next generation communications: Kickoff meeting. In Proc. DARPA (Vol. 10, p. 388).

\bibitem{b23} G. Zhao et al., "Spatial spectrum holes for cognitive radio with relay-assisted directional transmission," in IEEE Transactions on Wireless Communications, vol. 8, no. 10, pp. 5270-5279, October 2009, doi: 10.1109/TWC.2009.081541.

‌\bibitem{b24} G. Joshi, S. Nam, and S. Kim, “Cognitive Radio Wireless Sensor Networks: Applications, Challenges and Research Trends,” Sensors, vol. 13, no. 9, pp. 11196–11228, Aug. 2013, doi: https://doi.org/10.3390/s130911196.

\bibitem{b25} K.-C. Chen, Y. Peng, N. R. Prasad, Y.-C. Liang, and S. Sun, “Cognitive radio network architecture,” Jan. 2008, doi: https://doi.org/10.1145/1352793.1352817.

\bibitem{b26} H. Gao, C. Liu, Y. Li and X. Yang, "V2VR: Reliable Hybrid-Network-Oriented V2V Data Transmission and Routing Considering RSUs and Connectivity Probability," in IEEE Transactions on Intelligent Transportation Systems, vol. 22, no. 6, pp. 3533-3546, June 2021, doi: 10.1109/TITS.2020.2983835.

‌\bibitem{b27} R. M. Thomas, D. S. Friend, L. A. DaSilva, and A. B. MacKenzie, “Cognitive Networks,” pp. 17–41, Aug. 2007.

‌\bibitem{b28} J. K. Ray, A. Singh, Q. M. Alfred, S. Shome and R. Bera, "5G URLLC Communication System With Cognitive Radio and Frequency Diversity Reception For Improving Reliability In Smart Factory E-cranes operation," 2019 IEEE MTT-S International Microwave and RF Conference (IMARC), Mumbai, India, 2019, pp. 1-5, doi: 10.1109/IMaRC45935.2019.9118760.

\bibitem{b29} “Base Station Design and Architecture for Wireless Sensor Networks - The Robotics Institute Carnegie Mellon University,” www.ri.cmu.edu. https://www.ri.cmu.edu/publications/base-station-design-and-architecture-for-wireless-sensor-networks/ (accessed Jul. 30, 2023).

\bibitem{b30} R. Ali, D. N. Hakro, M. R. Tanweer and A. A. Kamboh, "Simulation based Vehicle to Vehicle and base station communication," 2019 International Conference on Information Science and Communication Technology (ICISCT), Karachi, Pakistan, 2019, pp. 1-6, doi: 10.1109/CISCT.2019.8777411.

\bibitem{b31} D. K. R and R. A, “Revolutionizing Intelligent Transportation Systems with Cellular Vehicle-to-Everything (C-V2X) technology: Current trends, use cases, emerging technologies, standardization bodies, industry analytics and future directions,” Vehicular Communications, vol. 43, p. 100638, Oct. 2023, doi: https://doi.org/10.1016/j.vehcom.2023.100638.
‌

\bibitem{b32} 3rd Generation Partnership Project; Technical Specification Group Services and System Aspects; Architecture enhancements for V2X services (Release 14), 3GPP TS 23.285, V14.1.0, Dec 2016.

\bibitem{b33} S. Gyawali, S. Xu, Y. Qian and R. Q. Hu, "Challenges and Solutions for Cellular Based V2X Communications," in IEEE Communications Surveys and Tutorials, vol. 23, no. 1, pp. 222-255, Firstquarter 2021, doi: 10.1109/COMST.2020.3029723.

‌\bibitem{b34} F. A. Teixeira, V. F. e Silva, J. L. Leoni, D. F. Macedo, and J. M. S. Nogueira, “Vehicular networks using the IEEE 802.11p standard: An experimental analysis,” Vehicular Communications, vol. 1, no. 2, pp. 91–96, Apr. 2014, doi: https://doi.org/10.1016/j.vehcom.2014.04.001.

‌\bibitem{b35} “5G; Study on channel model for frequencies from 0.5 to 100 GHz", 3GPP TR 38.901 version 15.0.0 Release 15.” Accessed: Jul. 30, 2023.

\bibitem{b36} D. Johnson, “Signal-to-noise ratio,” Scholarpedia, vol. 1, no. 12, p. 2088, 2006, doi: https://doi.org/10.4249/scholarpedia.2088.

‌\bibitem{b37} C. E. Shannon, "A mathematical theory of communication," in The Bell System Technical Journal, vol. 27, no. 3, pp. 379-423, July 1948, doi: 10.1002/j.1538-7305.1948.tb01338.x.

\bibitem{b38} Recommendation, I. T. U. T. "E. 800, Definitions of terms related to quality of service." International Telecommunication Union’s Telecommunication Standardization Sector (ITU-T) Std (2008).

‌\bibitem{b39} “5G; Service requirements for enhanced V2X scenarios (3GPP TS 22.186 version 16.2.0 Release 16).” Accessed: Jul. 30, 2023.

\bibitem{b40} M. Garg, C. Johnston and M. Bouroche, "Can Connected Autonomous Vehicles really improve mixed traffic efficiency in realistic scenarios?," 2021 IEEE International Intelligent Transportation Systems Conference (ITSC), Indianapolis, IN, USA, 2021, pp. 2011-2018, doi: 10.1109/ITSC48978.2021.9565068.

\bibitem{b41} T. Islam, Y. Hu, E. Onur, B. Boltjes and J. F. C. M. de Jongh, "Realistic simulation of IEEE 802.11p channel in mobile Vehicle to Vehicle communication," 2013 Conference on Microwave Techniques (COMITE), Pardubice, Czech Republic, 2013, pp. 156-161, doi: 10.1109/COMITE.2013.6545061.

\bibitem{b42} R. Molina-Masegosa, J. Gozalvez and M. Sepulcre, "Comparison of IEEE 802.11p and LTE-V2X: An Evaluation With Periodic and Aperiodic Messages of Constant and Variable Size," in IEEE Access, vol. 8, pp. 121526-121548, 2020, doi: 10.1109/ACCESS.2020.3007115.

\bibitem{b43} M. H. C. Garcia et al., "A Tutorial on 5G NR V2X Communications," in IEEE Communications Surveys and Tutorials, vol. 23, no. 3, pp. 1972-2026, thirdquarter 2021, doi: 10.1109/COMST.2021.3057017.

\bibitem{b44} N. G. Gupta, R. D. Thakre and Y. A. Suryawanshi, "VANET based prototype vehicles model for vehicle to vehicle communication," 2017 International conference of Electronics, Communication and Aerospace Technology (ICECA), Coimbatore, India, 2017, pp. 207-212, doi: 10.1109/ICECA.2017.8203672.



\end{thebibliography}
\end{document}